%% file: MWA_Pipelines.tex
\documentclass[twolcolumn,iop]{emulateapj}

\usepackage[backref,breaklinks,colorlinks,citecolor=blue]{hyperref}
\usepackage{graphicx}
\usepackage[space]{grffile}
\usepackage{latexsym}
\usepackage{amsfonts,amsmath,amssymb}
\usepackage{url}
\usepackage[utf8]{inputenc}
\usepackage{fancyref}

\hypersetup{pdfborder={0 0 0},}
\usepackage{longtable}
\usepackage{multirow,booktabs}
\usepackage[perpage]{footmisc}
\usepackage{natbib}
\usepackage{mathrsfs}

\shorttitle{ MWA Power Spectra}
\shortauthors{D. Jacobs}

\def\eppsilon{{\it $\varepsilon$ppsilon}}
\def\empirical{EmpCov}
\def\chipscite{\cite{2016arXiv160102073T}}
\def\eppsiloncite{Hazelton et al.\ 2016, \emph{in prep}}
\def\dilloncite{\cite{PhysRevD.91.123011} }

\begin{document}


\title{The Murchison Widefield Array 21 cm Power Spectrum Analysis Methodology}

%% Author list
\include{mwa_eor_collab}

\begin{abstract}
We present the 21 cm power spectrum analysis approach of the Murchison Widefield Array Epoch of Reionization project.  In this paper, we compare the outputs of multiple pipelines for the purpose of validating statistical limits cosmological hydrogen at redshifts between 6 and 12. Multiple, independent, data calibration and reduction pipelines are used to make power spectrum limits on a fiducial night of data.  Comparing the outputs of imaging and power spectrum stages highlights differences in calibration, foreground subtraction and power spectrum calculation. The power spectra found using these different methods span a space defined by the various tradeoffs between speed, accuracy, and systematic control.  Lessons learned from comparing the pipelines range from the algorithmic to the prosaically mundane; all demonstrate the many pitfalls of neglecting reproducibility. We briefly discuss the way these different methods attempt to handle the question of evaluating a significant detection in the presence of foregrounds.

\end{abstract}


\bibliographystyle{apj_w_etal}

\section{Introduction} 
  Study of primordial hydrogen  in the early universe via 21\,cm radiation has been forecast to provide a wealth of astrophysical and cosmological information.   Hydrogen is the principal product of big bang nucleosynthesis and is neutral over cosmic time from recombination until re-ionized by the first batch of UV emitters. While neutral it is visible in the 21\,cm radio line, which is both optically thin and spectrally narrow, making possible full tomographic reconstruction of a very large fraction of the cosmological volume.  Reviews of 21 cm cosmology, astrophysics and observing can be found in \cite{Morales:2010p8093,Furlanetto:2006p2267,Pritchard:2012p9555,zaroubi2013epoch}.
  
Direct detection of HI during the Epoch of Reionization (cosmological redshifts $5<z<13$) is currently the goal of several new radio arrays. The LOw Frequency ARray \citep[LOFAR;][]{Yatawatta:2013p9699}, the Donald C. Backer Precision Array for Probing the Epoch of Reionization \citep[PAPER;][]{Parsons:2014p10499} and the Murchison Widefield Array (MWA; \cite{Tingay:2013p9022,Bowman:2013p9950}) are all currently conducting long observing campaigns.

The analysis of the resulting data presents several challenges. The signal is faint; initial detection is being sought in the power spectrum with thousands of hours of integration (accumulated over several years) required to make a statistical detection, most commonly in the power spectrum. This faint spectral line signal sits atop a continuum foreground four orders of magnitude brighter \citep{Santos:2006p6697,Bowman:2009p7816,Pober:2013p9942}. At the same time, the instruments are fully correlated phased arrays with wide fields of view that strain the conventional mathematical approximations of radio astronomy practice (e.g. \cite{2007TMS}). The methods used to arrive at a well-calibrated, foreground-free estimation of the power spectrum are all under development; both the algorithms as well as their implementation.

{\bf New tools for Power Spectra in the Presence of Foregrounds}\\

The path from observation to power spectrum can be roughly divided into two parts: removal of foregrounds and estimation of power spectrum. 
Methods for estimating the power spectrum, particularly those which minimize the effects of foregrounds, have been studied and implemented by \citet{Morales:2006p1870,Jelic:2008p2130,Harker:2009MNRAS.397.1138H,Morales:2012p8790,Liu:2011p8763,Trott:2012p10466,Chapman:7p8505,Chapman:2013p10379,Dillon:2013p10497,Dillon:2014p9788,2014PhRvD..90b3018L,2014PhRvD..90b3019L}. Common elements include using knowledge about the instrument and foregrounds to minimize and isolate foreground contamination, applying the  quadratic estimator formalism to achieve favorable error properties, and studies of effects related to including the spectral dimension in the Fourier transform; i.e. translating two dimensional power spectrum techniques from CMB applications into three dimensional which also take a Fourier transform along the spectral/line of sight dimension.   One significant problem studied has been minimizing the impact of any residual foregrounds by downweighting or minimizing correlation with contaminated band powers. In this paper we compare power spectra calculated using a range of methods.

 The various dimensions of the 3D power spectrum space are used frequently throughout this paper.  Transverse modes $k_\perp$ are directly sampled by baselines of length $|u|$\footnote{Note that the mapping between $k_\perp$ and baseline vector $u$ is only strictly true in the small field of view limit.} while line of sight modes $k_\parallel$ are measured as $\eta$ modes which are the Fourier dual to frequency. Note also that for shorter baselines there is an approximate equivalence between $\eta$ and the geometric delay of wavefronts moving across the array. Here we follow the conventions of \cite{Furlanetto:2006p2267} relating the measured modes to their cosmological counterparts using  a $\Lambda$CDM cosmology with $H_0=100h$ kms$^{-1}$Mpc$^{-1}, \Omega_M=0.27,\Omega_k=0, \Omega_k=0.73$ \citep{2013ApJS..208...19H_wmap9_parameters}.

The 21\,cm power spectrum increases steeply with decreasing wave number making it desirable to remove foregrounds on the largest spatial and spectral scales possible.   These foregrounds, by virtue of the correlation function of the interferometer, have a chromatic response in the instrument which has a spectral period that increases with baseline length and distance from phase center. In a 2D power spectrum spanning line of sight and angular modes this produces a wedge-shaped feature which has been much discussed in the literature \cite{Datta:2010p8781,Vedantham:2012p10297,Parsons:2012p8896,Pober:2013p9942,Morales:2012p8790,2014PhRvD..90b3018L,2014PhRvD..90b3019L,2015PhRvD..91b3002D,Trott:2012p10466,2015ApJ...804...14T}. Sources far from the phase center---near the Earth's horizon---show up in the topmost part of the wedge and thus contribute most strongly to nominally uncontaminated modes.

 Recently, two sorts of foreground removal have been suggested: methods which exploit detailed knowledge of foregrounds and those which are relatively agnostic about the individual foreground sources. Among the latter, several authors have described methods for fitting and removing smooth spectrum foregrounds from image cubes  \citep{Morales:2006p1903,Bowman:2009p7816,Liu:2009p4762,Liu:2011p8763,2012MNRAS.Chapman.423.2518C,Chapman:2013p10379,Dillon:2013p10497,Yatawatta:2013p9699}. These methods have been demonstrated to robustly remove foregrounds near the field center but are less effective for sources far from the central lobe of the primary beam, i.e. in the wedge \citep{2015ApJ...804...14T,2015ApJ...807L..28T,Pober:2016ApJ...819....8P}.  A second class of agnostic methods is the delay/fringe-rate filtering approach \citep{Parsons:2012p8896,2014PhRvD..90b3018L,2014PhRvD..90b3019L}, which has been applied to data from PAPER \citep{Parsons:2014p10499,	2015ApJ...809...61A,2015ApJ...801...51J}.  Applying time and frequency domain filters to the time ordered data, this technique uses a small amount of knowledge about the instrument to filter out the wedge at high dynamic range.  This method removes smooth spectrum foregrounds across the entire sky and is comparatively robust in the face of uncertainty about the instrument at the cost of losing some sensitivity.  
 
 Meanwhile, full forward modeling and subtraction of a sky model such as that implemented for LOFAR (see e.g. \cite{Jelic:2008p2130,Yatawatta:2013p9699}) has the goal of directly subtracting the sources responsible for the wedge and accessing the shortest, brightest 21\,cm wavemodes. To this requires a much higher fidelity model of the instrument and foregrounds across the entire sky, horizon to horizon \citep{2015ApJ...804...14T}.

The MWA foreground removal approach leverages the array's optimization for imaging to directly subtract known foregrounds in addition to the full range of treatments of residual foregrounds, including foreground avoidance and foreground suppression.  If successful, direct subtraction opens the most sensitive power spectrum modes, substantially improving the ability of early measurements to distinguish between reionization models \citep{Beardsley:2013p9952,Pober:2014p10390}. Recent work towards the goal of foreground subtraction includes better algorithmic handling of wide field imaging effects \citep{Tasse:2012p9459,Bhatnagar..2013ApJ,Sullivan:2012p9457,Ord:2010p8442,2014MNRAS.444..606O}, and continually improving catalogs of sky emission \citep{deOliveiraCosta:2008p2242,Jacobs:2011p8438,2013ApJ...776..108J,Hurley-walker:2014p45}. Ongoing operation of the first generation low frequency arrays---LOFAR, PAPER and MWA are all in their  third or fourth year of operation---continues to push the refinement of instrumental models (e.g. the work of \cite{2015RaSc...50..614N} in mapping the primary beam with satellites) and improve the accuracy of model subtraction.  At the same time, more complete surveys of foregrounds are currently under way. These include the MWA GLEAM\footnote{GLEAM: GaLactic and Extragalactic All-sky MWA} survey \citep{2015PASA...32...25W}, GMRT TGSS\citep{Intema:2016arXiv160304368I}  and the LOFAR MSSS\footnote{MSSS: Multi-frequency Snapshot Sky Survey} \citep{2015A&A...582A.123H}..

In turn, efforts with these currently operational experiments are having a major influence on how future, larger, EoR experiments will be designed and conducted.  Primary among these future experiments will be programs using the low frequency Square Kilometre Array (\cite{2014aska.confE...1K}) and the Hydrogen Epoch of Reionization Array \citep[HERA;][]{Pober:2014p10390}.  Specifically, the MWA is one of three official precursor telescopes for the SKA and the only one of the three fully operational for science.  The low frequency SKA will be located at the MWA site in Western Australia, giving the MWA special significance.

Given the challenges of using newly developed methods to reduce data from a novel instrument to make a low sensitivity detection, it is reasonable to consider the question of how one knows one is getting the ``right'' answer.  One option is to generate, as accurately as possible, a detailed simulation of the interferometer output and then input that to the pipeline under test.  Such forward modeling is an essential tool for checking correct operation of portions of the pipeline; however, the model will always be an imperfect reflection of reality, leaving open multiple interpretations of any differences between model and data.  Forward modeling the instrument response is also difficult to divorce from the analysis pipeline being tested; often the same software doing the analysis is used to perform the simulations. A second option, and the focus of this paper, is comparison between multiple independent pipelines.

In section \ref{sec:observing} we summarize the observing strategy used to collect our data, section \ref{sec:pipelines} explains our multiple pipelines and comparison strategy. In section \ref{sec:results} we show comparisons of images, diagnostic power spectra, power spectrum limits, section \ref{sec:lessons} lists some lessons learned from the comparison process and section \ref{sec:conclusion} offers some conclusions.

\section{Observing}
\label{sec:observing}
%The MWA
\subsection{The MWA}
The MWA is an interferometric array of phased array tiles operating in the 80-300\,MHz radio band. Each tile consists of a 4x4 grid of dual polarization bow-tie shaped dipoles that are used to form a beam on the sky with a full width of 26\arcdeg($\lambda/2$m) at the half power point. Signals from individual antennas are summed by an analog delay-line beamformer which can steer the beam in steps of 6.8\arcdeg$cos(\textrm{elevation})$.  The signal is digitized over the entire bandwidth but only 30.72\,MHz are available at any one time.  This 30\,MHz of bandwidth is broken into 1.28\,MHz ``coarse'' bands by a polyphase filter-bank in the field and sent to the correlator \citep{Ord:2015PASA...32....6O} where it is further channelized to 40kHz, cross-multiplied and then averaged at 0.5 second intervals.  More details on the design and operation of the MWA can be found in \cite{Lonsdale:2009p7913} and \cite{Tingay:2013p9022}.

\subsection{The 21\,cm Observing Program}
\begin{figure*}[htbp]
\begin{center}
\includegraphics[width=\textwidth]{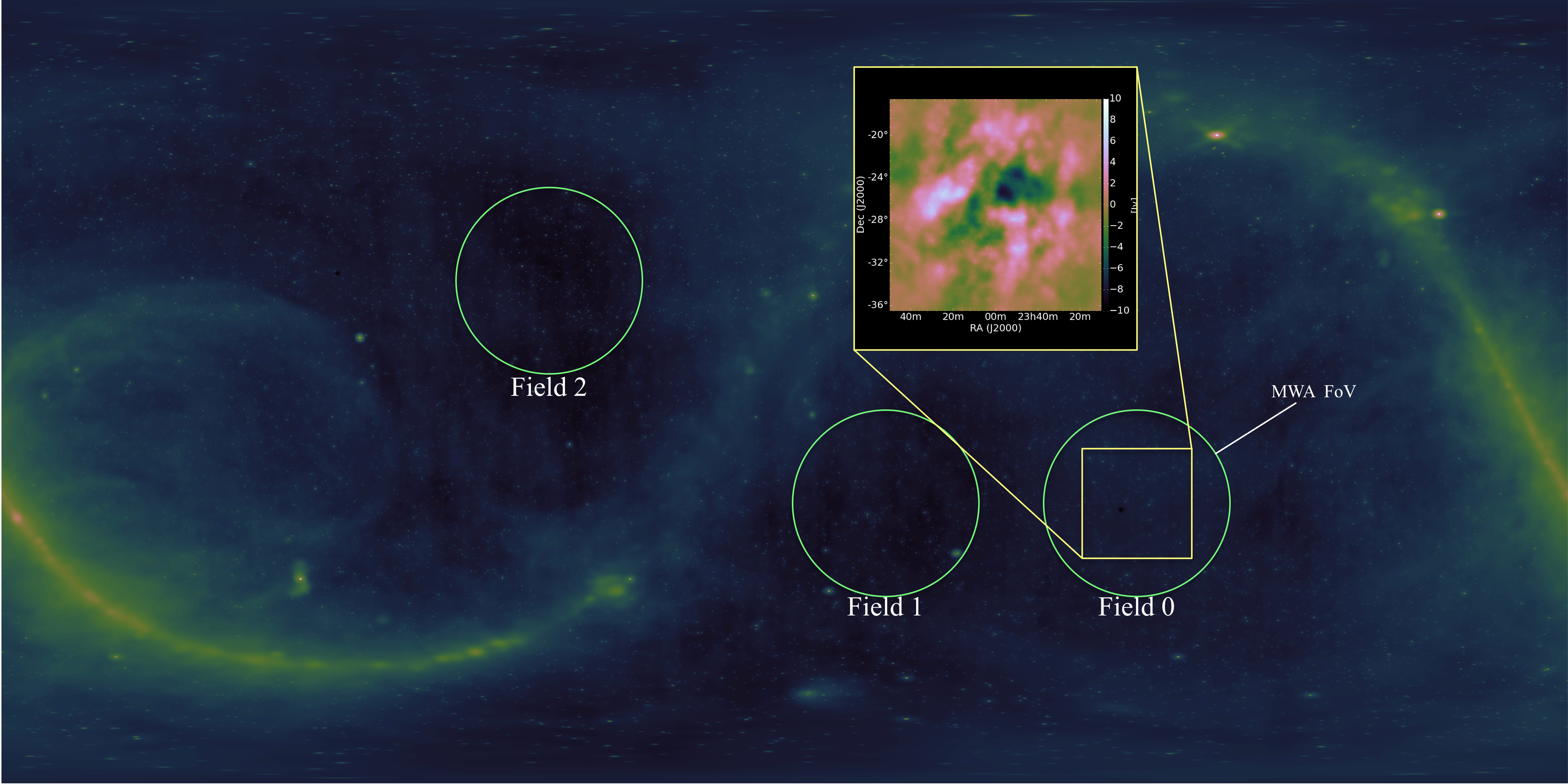}
\caption{An overview of the MWA reionization observation strategy. The background image is a cartesian view of the sky at radio wavelengths\footnote{Visualization compiled from NVSS \citep{Condon:1998p7986}, SUMSS \citep{Mauch:2003p8804}, and \citep{deOliveiraCosta:2008p2242}} and the circles indicate the deep fields observed by the MWA EoR project.  Here we are focusing on field 0, centered on Dec -27\arcdeg{} and RA 0h. Inset is a foreground subtracted image of the field made using the Real Time System (described more completely in \ref{sec:RTS}). A model of smooth (Galactic and unresolved) emission has not been subtracted and dominates the residual map of this 22\arcdeg{} wide image. Not visible in this average map is variation from channel to channel caused by sources far beyond the field of view which shows up as the wedge in 2d power spectra.}
\label{fig:fields}
\end{center}
\end{figure*}

The MWA reionization observing scheme spans two 30\,MHz tunings, 140-170\,MHz (9.2$<z<$7.5) and 167-196\,MHz (7.5$<z<$6.25) and two primary minimal foreground regions (Field 0 at RA 0h,-27\arcdeg and Field 1 at 4h, Dec -27\arcdeg); both transit the zenith at the MWA's  latitude and are near the south galactic pole. A third pointing towards Hydra A is also observed; see Figure \ref{fig:fields} for an overview. Here we focus on the low redshift tuning, and the RA=0h pointing, where the band is chosen for its lower sky temperature and pointing is chosen for its ease of calibration---having fewer bright, resolved sources; see Table \ref{tab:observing} for a listing of observing parameters.

The analysis presented here is on three hours of data, one of 400 nights which have been collected as part of the observing program; 150 nights are thought to be necessary for a detection of typical models \citep{Beardsley:2013p9952}.

%The data included here
\subsection{Data Included Here}
During observing, the beam-former was set such that the target region repeatedly drifted through the field of view.  With an available beamformer step size of 6.8\arcdeg; each drift was about 30 minutes long.  This was done for a total of 6 pointings in a night, or about 3 hours. The data included here include the two pointings leading up to the target crossing zenith, the zenith pointing, and then three more pointings after the transit crossing.  Data were recorded in 112 second units (which we term ``snapshots'') for a total of 96 snapshots. These snapshots are the basic unit of time on which many operations become independent -eg RFI flagging, calibration and imaging.   Though the full set of linear polarization parameters are correlated, and Stokes I images and power spectra are the final product of interest, at this stage of the analysis the instrumental polarizations have been found to be more instructive; with one exception, only the linear east-west polarization is examined here. No significant differences are seen in the north-south data.  The same set of snapshots is used in every pipeline run.

\subsection{Interference Flagging}
Each snapshot is flagged for interference using the AOFlagger \citep{offringa:2010rfim.workE..36O}\footnote{ \url{sourceforge.net/projects/aoflagger} } algorithm which applies a high pass filter to remove foregrounds, uses the SumThreshold algorithm to search for line-shaped outliers and then applies a scale invariant rank operator to the resulting mask which identifies certain morphological features.  

 As described in \cite{2015PASA...32....8O}, the interference environment at the Murchison Radio-astronomy Observatory is relatively benign and generally requires flagging of about 1.1\% of the data.  After flagging the data are averaged to 2 seconds and 80kHz, reducing the data volume by a factor of 8. 
\begin{deluxetable*}{lcr}
\tablecolumns{2}
\tablecaption{MWA EoR Observing Parameters }
\tablehead{
\colhead{Parameter}  & 
\colhead{Value} 
}
\startdata
Field of View & 26\arcdeg$(\lambda/2)$m FWHM \tabularnewline
Tuning & 166-196\,MHz  redshift range $7.56<z<6.25$ \tabularnewline
Target area & (RA,Dec) 0h00m, -27\arcdeg00m \tabularnewline
Primary beam pointing quantization & 6.8\arcdeg \tabularnewline
Snapshot length & 112 seconds\tabularnewline
Time and frequency resolution & 0.5\,s, 40 kHz  \tabularnewline
Post-flagging resolution & 2s, 80\,kHz \tabularnewline
Time & 3 hours on August 23, 2013, six 30 minute pointings or 96 snapshots\tablenotemark{a} 
\tabularnewline
\enddata
\tablenotetext{a}{The same data set is used in every pipeline run}
\label{tab:observing}
\end{deluxetable*}

\section{Power Spectrum Pipelines}
\label{sec:pipelines}
In this section we introduce the basic pipeline components, define some terms common to all, and then in sections \ref{sec:RTS}-\ref{sec:CHIPS} give finer grain descriptions of the specific implementations.

%about the power spectrum
The 21\,cm brightness at high redshift is weak and detectable by first generation instruments only in	 statistical measures such as the power spectrum. The spectral line signal is a three dimensional probe, two spatial dimensions and a third from the mapping of the spectral axis to line-of-sight distance via the Hubble relation. 3D power spectra are computed at multiple redshift slices through the observed band and then, taking advantage of the cosmological signal's statistical isotropy, averaged in shells of constant wavenumber $k$.  The power spectrum is well matched to an interferometer, which natively measures spatial correlation; the baseline vector maps to the perpendicular wavemode $k_\perp$.  An additional Fourier transform in the spectral dimension provides $k_\parallel$.  

%about foregrounds
The principal challenge to detecting 21\,cm at very high redshifts is foreground emission. At frequencies below 200\,MHz the dominant sources are synchrotron emissions from the local and extragalactic sources. Synchrotron is generally characterized by a smooth spectrum which rises as a power law towards lower frequencies. The local Galactic neighborhood has a significant amount of spatially smooth power appearing at short $k_\perp$ modes; extragalactic point sources appear equally on all angular scales and dominate over the Galaxy on long $k_\perp$ modes.

Our analysis pipeline has two main components: one which removes foregrounds--leaving as small a residual as possible---and a second which computes an estimate of the power spectrum.   Foreground subtraction is generally the domain of calibration and imaging software where the focus is on building an accurate forward model of the telescope and foregrounds.  Challenges include: ionospheric distortion, a very wide field of view, primary beam uncertainty, polarization leakage, and catalog inaccuracy. Though a number of calibration and imaging software packages---such as CASA and Miriad---are available, these challenges have necessitated the creation of custom software.  As an added benefit, having developmental control of the imager enables the export of additional information describing the observation which are necessary for the calculation of power spectrum uncertainties as described below. 

As we mentioned in the introduction, a horizon-to-horizon model of the sky must be subtracted at high precision from each 112s snapshot across thousands of hours of data. At this scale, deconvolution and self-calibration of each snapshot image is not computationally tractable.  In both FHD and RTS, with the exception of a small number of peeled sources, the sky model is not a part of the fit; rather than peeling a large number of sources the focus has been on refining the instrument model used to subtract catalogs.  This instrument model also provides information on the instrumental covariance which is used by the power spectrum estimators.

Detailed knowledge of instrumental covariance is essential to overcoming the two main challenges in estimating the power spectrum: 1) minimizing the effects of residual foregrounds and 2) faithfully recovering the underlying 21\,cm power.   As discussed in the introduction, simulations and early observations have shown that foregrounds tend to contaminate only specific $k$ modes; using a model of instrumental covariance the power can be isolated to fewer modes.  Accurate recovery of the 21\,cm background will, to first order, depend on the ability to correctly calculate uncertainties.  Initial power spectra are expected to be of low signal to noise \citep{Pober:2014p10390,Beardsley:2013p9952}, an accurate estimate of error is essential to estimating the significance of any putative detection .

\begin{figure*}[htbp]
\begin{center}
\includegraphics[width=\textwidth]{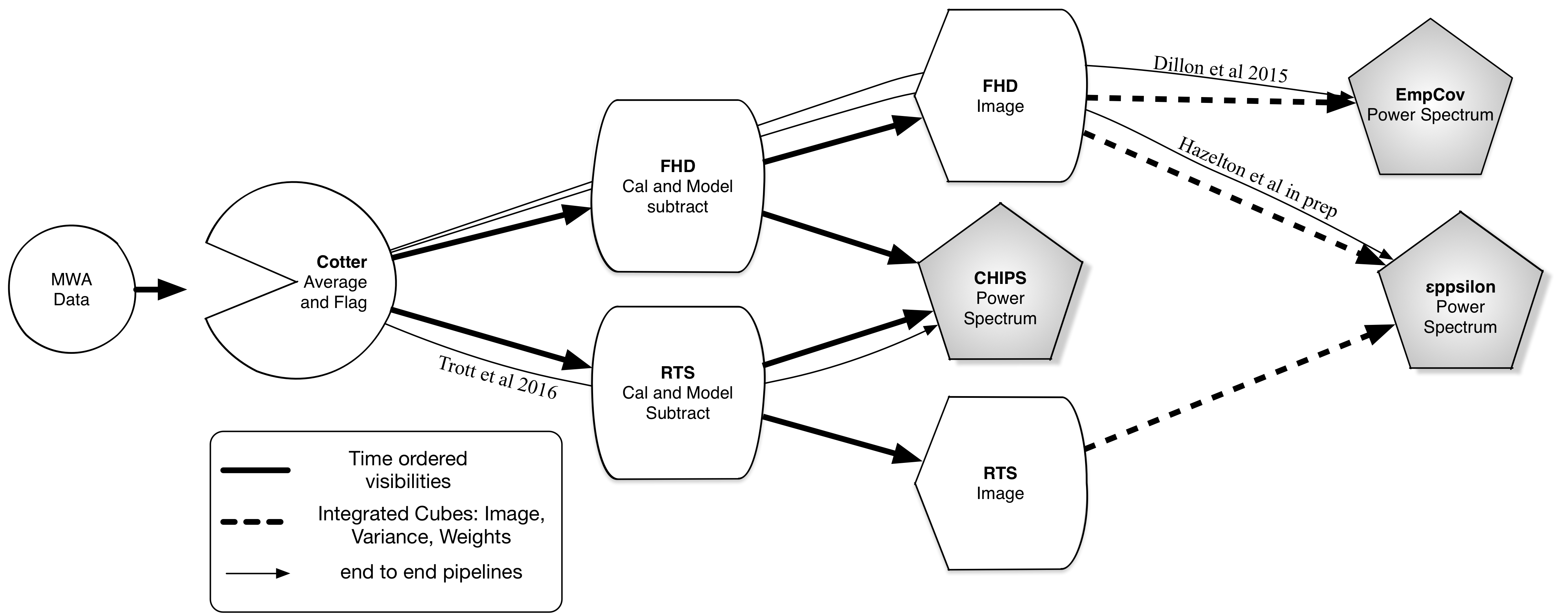}
\caption{Parallel pipelines with cross-connections after foreground subtraction and imaging are compared against each other. Pipelines used to reach the cited power spectrum results are indicated with thin lines; citations for each block are listed in Table \ref{tab:pipeline_cites}. Cotter uses AOFlagger to flag RFI and averages by a factor of 8. The averaged data are passed to either FHD or RTS for calibration, foreground subtraction and imaging. Both of these packages generate integrated residual spectral image cubes as well as matching cubes of weights and variances.  \eppsilon{} and \empirical{} use these cubes to estimate the power spectrum. Meanwhile, CHIPS taps into the RTS and FHD data stream to get calibrated and foreground-subtracted time-ordered  visibilities which it then grids with its own instrument model to estimate the power spectrum. 
}
\label{fig:pipes}
\end{center}
\end{figure*}

  Within the MWA collaboration, efforts have centered around multiple independent paths from raw data to a power spectrum.  As described in Figure \ref{fig:pipes}, these pipelines are generally divided into a component which performs calibration, foreground subtraction and imaging, and one which computes the power spectrum.  During development, each power spectrum code was paired with a ``primary'' foreground subtraction method, FHD with \eppsilon{} and RTS with CHIPS.   The main results come from these primary paths (as depicted by the thin lines in Figure \ref{fig:pipes}).  A third power spectrum estimator, \empirical{} by \dilloncite{}, has also been connected to the FHD imager. 

The primary difference between these pipelines is the division of responsibilities between foreground subtraction and power spectrum calculation. Some power spectrum methods take as input spectral image cubes output by the calibration and foreground subtraction system.  These image-based methods use a model of the instrument to inverse variance weight the data as it is averaged from the time domain into an image cube, the quadratic sum of the weights is also recorded to enable propagation of the weights into error bars on the averaged power spectrum.  
  Each set of cubes is generated by including every other integration at both even and odd sample cadences; the cross multiplication provides a power spectrum free of noise bias and the difference is an estimate of noise.

Methods which take time-ordered data as input generate their own instrument model internally.  The pipeline submodules names and citations are listed in Table \ref{tab:pipeline_cites} and described individually in sections \ref{sec:RTS} - \ref{sec:empirical_cov}.  

\begin{deluxetable*}{llr}
\tabletypesize{\footnotesize}
\tablecolumns{2}
\tablecaption{MWA EoR Pipeline components }
\tablehead{
\colhead{Short Name} &
\colhead{Name}  & 
\colhead{Citations} 
}
\startdata
Cotter & AOFlagger + Averaging & \cite{offringa:2010rfim.workE..36O} \tabularnewline
RTS & Real Time System&\cite{Mitchell:2008p707,Ord:2010p8442} \tabularnewline
FHD & Fast Holographic Deconvolution &\cite{Sullivan:2012p9457}\tablenotemark{1}  \tabularnewline
\eppsilon{} & Error Propagated Power Spectrum with InterLeaved Observed Noise & \eppsiloncite{}\tablenotemark{2} \tabularnewline
CHIPS & Cosmological HI Power Spectrum& \chipscite{}  \tabularnewline
\empirical{} & Empirical Covariance Estimator & \dilloncite{}

\enddata
\tablenotetext{1}{\url{github.com/miguelfmorales/FHD}}
\tablenotetext{2}{\url{github.com/miguelfmorales/eppsilon}}
\label{tab:pipeline_cites}
\end{deluxetable*}

\subsection{Calibration and Imager \#1: RTS}
\label{sec:RTS}

The MWA Real Time System (RTS; \cite{Mitchell:2008p707,Ord:2010p8442}) was initially designed to make wide-field images in real time from the MWA 512-tile system \citep{Mitchell:2008p707}.  On the de-scoped 128 element array, it has been implemented as an offline system, where it has been adjusted to compensate for the lower filling factor \citep{Ord:2010p8442}.  The RTS incorporates algorithms intended to address a number of known challenges inherent to processing MWA data, including; wide-field imaging effects, direction-dependent (DD) antenna gains and polarization response, and ionospheric refraction of low-frequency radio waves. Each MWA observation (112s) is processed by the RTS separately, in series. The RTS is also parallelized over frequency so that each coarse channel (1.28\,MHz broken into 40 kHz channels) is processed largely independently of the other coarse channels, with only information about peeled source offsets communicated between processing nodes.   

The RTS calibration strategy is based upon the `peeling' technique proposed by \cite{Noordam:2004p2379} and a foreground model using a cross-matching of heritage southern sky catalogs\footnote{See Table \ref{tab:cal_sub_parms}} with the MWA Commissioning Survey\footnote{The cross-matching is done using the PUMA code (Line et al, in prep) which uses Bayesian inference to build a self-consistent set of SEDs for sources using data from catalogs with varying frequency and resolution}.   The brightest apparent calibrators in the field of view are sequentially and iteratively processed through a Calibrator Measurement Loop (CML). During each pass through the CML; i) the expected (model) visibilities of known catalog sources are subtracted from the observed visibilities. For the data processed in this work, 1000 sources are subtracted for each observation. ii) The model visibilites for the targeted source are added back in and phased to the catalog source location. Any ionospheric offset of the source can now be measured by fitting a phase ramp to the phased visibilities. iii) The strongest sources are now used to update the direction-dependent antenna gain terms, while weaker sources are only corrected for ionospheric offsets. For this work, 5 sources are used as full DD calibrators and 1000 sources are set as ionospheric calibrators. The CML is repeated until the gain and ionospheric fits converge to stable values. A single bandpass for each tile is found by fitting a 2nd order polynomial to each 1.28\,MHz-wide coarse channel. The $\sim$1000 strongest sources are then subtracted from the calibrated visibilities.   Calibration and model subtraction parameters are summarized in Table \ref{tab:cal_sub_parms}.  Model subtracted visibilities are passed to the RTS imager and to the CHIPS power spectrum estimator. 

The RTS imager uses a snapshot imaging approach to mitigate wide-field and direction-dependent polarization effects. Following calibration, the residual visibilities are first gridded to form instrumental polarization images which are co-planar with the array. These images are then regridded into the HEALpix \citep{Gorski:2005p7667} frame with wide-field corrections.  Weighted instrument polarization images are stored, along with weight images containing the Mueller matrix terms, so that further integration can be done outside of the RTS. It is also possible to use the fitted ionospheric calibrator offsets to apply a correction for ionospheric effects across the field during the regridding step or subtraction of catalog sources, but in this work this correction has not been applied. These snapshot data and weight cubes are then integrated in time to produce a single HEALpix cube. This cube, averaged over the spectrum, is shown, with and without foregrounds, in Figure \ref{fig:image_compare}.

\begin{deluxetable}{lll}
\tablecolumns{3}
\tabletypesize{\footnotesize}
\tablewidth{0pt} 
\tablecaption{MWA Reionization Calibration and Model subtraction Parameters. Counts are per-snapshot unless otherwise noted }
\tablehead{
\colhead{Parameter}  & 
\colhead{RTS} &
\colhead{FHD} 
}
\startdata
per cable passband & NA & 384 channels\tablenotemark{a}   \\
per antenna passband & 48 per tile\tablenotemark{b} & 3\tablenotemark{c}\\
per antenna gain & 2\tablenotemark{d} & 2\tablenotemark{d}  \\
peeling parameters & 4\tablenotemark{e} & None \\
peeled sources & 5 & None\\
subtraction catalog & Line\tablenotemark{f} & Carroll\tablenotemark{g} \\
number subtracted & 1000 & 6932 \\
\bf{Total free parameters} & \bf{6,420} & \bf{880} \\
\enddata
\tablenotetext{a}{for ea. of 6 cable types, averaged over 96 snapshots}
\tablenotetext{b}{2nd order polynomial per coarse channel}
\tablenotetext{c}{poly fit over full band, 2nd order for amp, 1st for phase}
\tablenotetext{d}{amplitude and phase}
\tablenotetext{e}{Direction Dependent (DD) gain fits}
\tablenotetext{f}{MWA Commissioning Survey\cite{Hurley-walker:2014p45},\cite[VLSSr]{2014MNRAS.440..327L},\cite[MRC]{Large:1991p7760},\cite[SUMSS]{Mauch:2003p8804},\cite[NVSS]{Condon:1998p7986},cross matched using PUMA (Line et al, in prep) \url{github.com/JLBLine/PUMA}}
\tablenotetext{g}{Combination catalog of legacy catalogs and sources deconvolved from this data (Carroll et al in prep)}
\label{tab:cal_sub_parms}
\end{deluxetable}

\subsection{Calibration and Imager \#2: FHD}
\label{sec:FHD}
Fast Holographic Deconvolution (FHD, \cite{Sullivan:2012p9457}) is an imaging algorithm designed for very wide field of view interferometers with direction- and antenna-dependent beam patterns. Simulated beam patterns are used to grid visibilities to the $uv$ plane and the reverse operation of turning gridded cubes into time-ordered visibilities. This simulation of weights provides a necessary accounting of information loss caused by the inherent size of the dipole element. As it is designed from the ground up to account for the widefield effects inherent in the MWA it has grown to include a full range of tasks such as calibration and simulation. In this pipeline the framework is used to both calibrate and image the data.

The FHD calibration pipeline generates a model data set, computes a calibration solution which minimizes the difference with the data, smooths the calibration solution to minimize the number of free parameters, and outputs the residual. The calibration model is formed from sources found by deconvolving, in broadband images, about 75 of the 96 snapshots included here and retains those which are common to all snapshots and pass other consistency checks (Carroll et. al. in prep). In each snapshot sources are included in the model if they are at or above 1\% of the peak primary beam, this amounts to about 7000 sources and a flux limit of about 80mJy with slight variations snapshot to snapshot. Most sources in this catalog have spectral indexes between 0 and -2 with the majority near a mean of $-0.8$, which corresponds to a 13\% difference across the 30\,MHz.  Spectral index is not directly modeled, spectra are simulated to be flat, however, during catalog subtraction, the data are multiplied by a positive spectral index of 0.8 such that most sources will appear to be flat. Full spectral modeling of catalog sources will be included in future analyses.

The function of FHD's calibration is to minimize the number of free parameters with the twin goals of minimizing potential signal loss and in building a deep understanding of instrumental systematics. Here we give a brief description of the instrument model; a listing of all the parameters mentioned here is also given in Table \ref{tab:cal_sub_parms} along with a rough accounting for the total number of fitting parameters. Initial complex gain solutions are computed using the Alternating Direction Implicit technique described in \citet{sal14} for each antenna, channel and polarization.  This generates a gain and phase for every channel on every tile, for each 112s snapshot.  Most antennas have similar solutions with the main features corresponding to the exact type and length of analog cable feed of which there are 6 different types; per-tile solutions are further averaged into per-cable-type and averaged over the entire 3 hour observation. After these solutions are divided out, the residual per-tile solutions are further fit for a second order amplitude spectral polynomial and a 1st order phase slope. This is done on every snapshot to account for temperature-driven amplifier gain changes. 
One systematic easily visible in the power spectra is a small reflection corresponding to the 150m cables. This is fit and removed as a phase delay with a $\sim$0.1dB amplitude in the time averaged per-tile bandpass solutions.

The residual time-ordered visibilites are then passed to CHIPS and to FHD imaging for formation of spectral cubes.  

{\bf FHD Imaging}\\
\label{sec:FHD_imaging}
The imager portion of the FHD framework is the Fast Holographic Deconvolution algorithm which is based on loss-less optimal map-making developed for the CMB \cite{1997ApJ...480L..87T}. It is a variant on the class of wide-field imaging algorithms such as  A-projection \citep{Bhatnagar..2013ApJ}, peeling \citep{Mitchell:2008p707}, and forward modeling \citep{Pindor:2011p10350,Bernardi:2011p9205} which improve upon standard radio interferometric algorithms by including the primary beam in the calculation of the instrumental response.  It is Holographic in the sense that it uses the known primary beam of the antenna when performing simulation and gridding operations, and Fast due to its use of a sparse matrix representation of the instrument transfer function.  
FHD is described in detail by \cite{Sullivan:2012p9457}.  Deconvolution is not used in this pipeline paper, however FHD's relevant focus on highly accurate primary beam modeling is used to form optimally weighted images. FHD does not currently grid $w$ terms separately and is therefore limited to planar arrays and relatively short periods of rotation synthesis where long baselines do not rotate significantly with the Earth.

As deployed in the pipeline described here, the FHD imager is used to grid the time ordered visibilities into a $uvf$ cube weighted according to the Fourier transform of the primary beam using an electromagnetic model described by \cite{Sutinjo:2015RaSc...50...52S}. This weighting scheme is similar in effect to ``natural'' weighting scheme which provides the highest possible SNR by weighting $uv$ samples according to the inverse variance. Natural weighting is traditionally not favored in interferometric imaging with sparse arrays as it can dramatically impact the point spread function by downweighting baselines for which there are few samples.  We choose it here as it maximizes the sensitivity of our densely sampled uv plane core. 

Deep mosaics are made by using the warped snapshot method \citep{2012SPIE.8500E..0LC}.  The data are split into even and odd samplings at the 8s cadence and then imaged at a two minute cadence with a $uv$ resolution chosen to result in a 90 degree field of view.  The even odd split is carried through to the final power spectrum analysis where they are cross-multiplied.  In each $uv$-frequency ($uvf$) voxel we accumulate weighted data, weights and the square of the weights according to 

\begin{equation}
\label{eq:weight_accumulation}
\begin{split}
D_{uvf} &= \sum_i B_{i;uvf} V_{i;uvf}\\
W_{uvf} &= \sum_i B_{i;uvf}\\
\textrm{Var}_{uvf} &= \sum_i B_{i;uvf} B_{i;uvf}^*
\end{split}
\end{equation}
where $B_i$ is the Fourier transform of the cross power beam evaluated in that $uvf$ voxel at snapshot time $i$,$V_i$ is visibility at time $i$\footnote{We are not assuming Einstein notation; all sums are written explicitly}. Essentially $D$ is numerator of the mean performed in the mosaicing step with $W$ the normalizing denominator of the mean; the same is done for the error cube Var/$W^2$.\footnote{Note that $D$/$W$, the cube used to calculate power spectra, represents our best estimate of the power perceived by the instrument; an image made from this cube would still be attenuated by one factor of the primary beam.   Dividing out by this last factor in the image space would substantially increase the $uv$ correlation length and invalidate our assumption of $uvf$ diagonality. Because we do not divide out by this factor of the beam, we do not form Stokes cubes, which are only defined for images with a uniform flux scale. The remaining factor of the primary beam constitutes a volume term in the power spectrum that we account for in the normalization of the power spectrum using the same conventions as CHIPS (\chipscite{}).}  The cubes are Fourier transformed, corrected for the $w$ projection coordinate warping, and gridded into the HEALpix frame.

\begin{figure*}[htb]
\begin{center}
\includegraphics[width=1\textwidth]{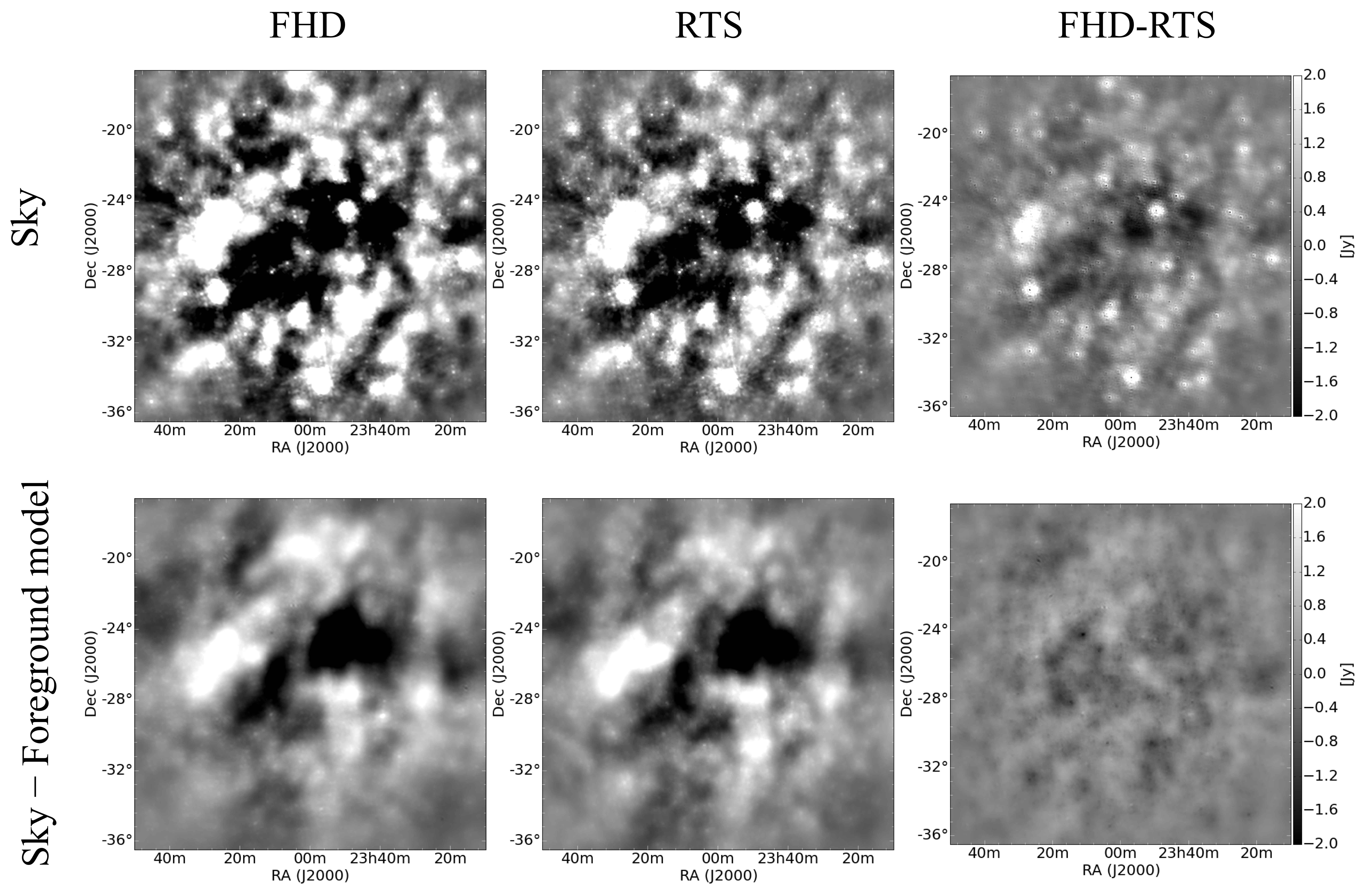}
\caption{A comparison between the image outputs of the FHD (left), RTS (center) and their difference (right) averaged in the spectral dimension and projected from native HEALpix to flat sky. The images have been left in the natural weighting used by image-based power spectrum schemes and no deconvolution has been applied.   In the top row, no foreground model has been subtracted; the residual shown represents a 15\% difference. On the bottom both have subtracted their best model of the sky containing similar sets of thousands of sources; in most pixels the difference is 30\% or lower.   The difference between foreground subtracted images reveals a good agreement on large scale structure and small differences in the fluxes of a few sources. Differences in these mean maps are very similar to the differences seen in the individual channels.
\label{fig:image_compare}}
\end{center}
\end{figure*}

{\bf Mosaicing}\\
 The 112 snapshot HEALpix cubes are summed in time, keeping pixels with a beam weight of 1\% or more, a cut which effectively limits the field of view to $\sim$20\arcdeg. The resulting mosaic is handed on to \eppsilon{} (S. \ref{sec:EPPSILON}) and  \empirical{} (S \ref{sec:empirical_cov}).  This image, averaged over the spectral dimension, is shown, with and without foregrounds, in Figure \ref{fig:image_compare}. This image retains the full weighting proportional to the number of samples in each $uvf$ cell and is therefore very similar to natural weighting. Though beam weighting theoretically gives an optimal inverse variance weighting for each snapshot it does not capture the change in variance due to changing system temperature. The mosaicing as performed here weights each snapshot equally, under the assumption that the system temperature, which is dominated by the temperature of the sky in the direction of the phased array pointing, stays roughly constant through the three hour tracked integration.

\subsection{Comparing Calibration and Imaging Steps}
\label{sec:comparing_imaging}
Through the parallel-but-convergent development of these imagers have emerged two very similar systems; however, some differences remain in the analysis captured here. The two primary differences are in the treatment of calibration and in the subtracted catalogs.  

In both pipelines the calibration is a two step process. First, calibration solutions for each channel, and antenna are computed by solving for the least-squares difference with a model data set. Next, those solutions are fit to a model of the array; for example fitting a polynomial to the bandpass. FHD and RTS take different approaches to this step, a fact reflected in the the number of free parameters in this fit. A smaller number of parameters minimizes the possibility of cosmological signal loss or the spurious incorporation of unmodeled foreground emission in the calibration solutions \citep{Barry:2016a}; more free parameters can absorb physics missing from the instrument model.  As tabulated in Table \ref{tab:cal_sub_parms} the RTS fits for $\sim$6,420 free parameters while FHD fits for $\sim$880.

In practice, some parameters will be averaged over more than a single night which will further reduce the number of free parameters per observation, though hundreds to thousands of free parameters is still typical. This is a large number but it is considerably smaller than the 180 million data points typically recorded in single snapshot obdservation. 

As has been noted, there is nearly an order of magnitude difference in the number of free parameters between the two pipelines, which is worth considering. The primary difference is in the treatment of the passband.  There are a number effects which show up in the passband calibration. The edges of the 1.28\,MHz bands are known to be subject to aliasing from adjacent coarse channels as well as under-sampling when cast to 4bit integers by the correlator (van Vleck corrections) and so are flagged. This flagging creates a regular sampling function which shows up as the characteristic horizontal lines in a 2D power spectrum (see section \ref{sec:power_spectrum_comparison}). Added to this is a small amount of interference flagging.  Additionally, reflections at analog cable junctions show up as additional spectral ripple corresponding to the length of the cables.    

The RTS fits for a low order polynomial on every 1.28\,MHz chunk on every antenna, while FHD averages each channel over all antennas to get a common passband for all and then fits a low order polynomial to get any tile to tile variation. This significantly reduces the number of free parameters and the likelihood of signal loss, though leaving open the possibility of additional un-modeled instrumental effects.

The construction of the foreground subtraction model is also a point of difference between the two pipelines.
As noted in Table \ref{tab:cal_sub_parms}, foreground/calibration models contain different numbers of sources which have been derived by different means. The RTS catalog cross-matches multiple heritage southern sky catalogs with the MWA Commissioning Survey using the Bayesian cross-matcher PUMA (Line et al in Prep).  The FHD subtraction model contains sources found in a deep deconvolution of this same data set. Both catalogs have the goal of producing a reliable set of sources that minimizes false positives and accurately reflects resolved components, though they go about it in different ways. The FHD catalog focuses on the reliability aspect by performing a deconvolution on every snapshot used in the observation and selecting sources which appear in most observations (Carroll et al, in prep). The RTS catalog has used the somewhat less precise MWA commissioning catalog but by cross-matching these sources against many other catalogs of known sources and fitting improved positions and fluxes, the accuracy is seen to increase.

\subsection{Power Spectrum \#1: \eppsilon}
\label{sec:EPPSILON}
\eppsilon{} calculates a power spectrum estimate from image cubes and
directly propagates errors through the full analysis, see \eppsiloncite{} for a full description. The design criteria for this method is to make a relatively quick and uncomplicated estimate of the power spectrum to provide a quick turnaround diagnostic.

The accumulated data, weight and variance cubes are Fourier transformed along the two spatial dimensions into $uvf$ space, where the spatial covariance matrix is assumed to be diagonal. This is approximately true if the $uv$ pixel size is well matched to the primary beam size, so the  \eppsilon{} direct Fourier transform grid size is restricted to being equal to the width of the Fourier transformed primary beam; i.e. 1/(field of view). The $uvf$ data cubes are then divided by the weight cubes to arrive at a uniformly weighted 3D Fourier cube. The variance cubes are similarly divided by the square of the weights. Next the sum and difference of the even and odd cubes are computed with variances given by adding the reciprocal of the even and odd variances in quadrature. The difference cube then contains only noise  and the sum cube contains both sky signal and noise.

The next step is to Fourier transform in the frequency direction. Here we choose to weight by a Blackman-Harris window function, which heavily downweights the outer half of the band and decreases the leakage of bright foreground modes into other power spectrum modes as described in \cite{Thyagarajan:2013p10039,Parsons:2012p8896,Vedantham:2012p9026}, among others.  The transform of the spectral dimension is done using the Lomb \& Scargle periodogram to minimize the effects of regular gaps in the spectrum which occur every 1.28\,MHz.    The sky signal power is  estimated by the square of the sum cube minus the square of the difference cube which  is mathematically identical to the even/odd cross power if the even and odd variances are identical

\begin{equation}
\begin{split}
((E + O) (E+O)^\dag &- (E-O)(E-O)^\dag)\frac{1}{4} = \\
 &= (EE^\dag + OO^\dag + EO^\dag + OE^\dag - \\
		& (EE^\dag+OO^\dag - EO^\dag - OE^\dag) )\frac{1}{4} \\
&= (2EO^\dag + 2OE^\dag)\frac{1}{4} \\
&=\Re(EO^\dag),
\end{split}
\end{equation}
while the square of the difference cube provides a realization of the noise power spectrum.

Diagnostic power spectra are generated by averaging cylindrically to a two dimensional $k_{\|},k_{\bot}$ power spectrum. The diagnostic power spectra are computed over the full 30\,MHz bandwidth to provide the highest possible $k_\parallel$ resolution of the foreground and systematic effects.  These are shown in the left column of Figure \ref{fig:pspec_compare}.  One dimensional power spectra (shown in Figure \ref{fig:1D_pspecs}), are calculated by a similar process but only using 10\,MHz\footnote{Hereafter,  unless otherwise noted, 2D and 1D power spectra from all pipelines will span 30\,MHz and 10\,MHz respectively.}  of bandwidth which corresponds to a redshift range of 0.3 (at 182\,MHz) and averaging along shells of constant $k$, masking points within the foreground wedge\footnote{Here defined, conservatively, as the light travel time across the baseline plus the delay associated with the pointing furthest from zenith. It is indicated as a solid line on Figure \ref{fig:pspec_compare}.} and weighting by inverse variance. 

The error bars are estimated as the sum of the beam weights accumulated in the mapping process (eg eq. \ref{eq:weight_accumulation}) assuming a constant noise figure for each data point

\begin{equation}
\sigma^2_{uvf} = \sum_i Var^i_{uvf}/(W^i_{uvf})^2  \sigma^2_{if}
\end{equation}
where the noise $\sigma^2_{if}$ is an average noise spectrum, per snapshot $i$, calculated by differencing even odd data sets and computing an rms over all baselines for each snapshot $i$. These errors are then propagated into the 2D and 1D power spectra by quadratic sum.

 As noted above, this noise model assumes that the apparent sky and receiver temperatures are roughly constant through the 3 hour tracked observing period. Another way of estimating error is to form power spectra from the even odd difference.  Comparing these two methods we find that they agree to within a few percent in both 2D and 1D power spectra.

\subsection{Power Spectrum \#2: CHIPS}
\label{sec:CHIPS}
The CHIPS power spectrum estimation method computes the maximum likelihood estimate of the 21~cm power spectrum using an optimal quadratic estimator formalism and is more completely described in \chipscite{}.  The design criteria for this method were to fully account for instrumental and foreground induced covariance in the estimation of the power spectrum.  The approach is similar to that used by \cite{Liu:2011p8763}, but with the key difference of being performed entirely in $uvw$-space, where the data covariance matrix is simpler (block diagonal), and feasible to invert. This approach also allows straightforward estimation of the variances and covariances between sky modes by direct propagation of errors. CHIPS takes as input calibrated and foreground subtracted time-ordered visibilities. Tapping into the pipeline post-calibration but before imaging, CHIPS uses its own internal instrument model to estimate and propagate uncertainty.	

The method involves four major steps: (1) Grid and weight time-ordered visibility channels onto a $uvw$-cube using the primary beam model, (2) compute the least squares spectral (LSS) transform along the frequency dimension to obtain the best estimate of the line-of-sight spatial sky modes (this technique is comparable to that used by \eppsilon), (3) compute the maximum-likelihood estimate of the power spectrum, incorporating foregrounds and radiometric noise,  averaging $k_x$ and $k_y$ modes into annular modes on the sky, $k_\bot$; (4) compute the uncertainties and covariances between power estimates. The first step is the most computationally-intensive, requiring processing of all the measured data. The main departure point for CHIPS from \eppsilon{} is in the much finer resolution of the $uv$ grid.  Using an instrument model, CHIPS calculates the covariance between $uv$ samples as a function of frequency.  Since the beam and $uv$ sampling function are both highly chromatic, extra precision in this inversion is thought to be highly beneficial. After a line of sight transform similar to that used by \eppsilon{}, this covariance information is inverted to find the Fisher Information, the maximum likelihood power spectrum, and covariances between measurements.  The maximum likelihood estimate of the power in each $k_\bot,k_\parallel$ mode is shown in the right column of Figure \ref{fig:pspec_compare} and averaged in spherical bins in Figure \ref{fig:1D_pspecs}. 

Before this last averaging step one can optionally include an additional weighting by the known power spectrum of a confused foreground in a process described in more detail for these data by \chipscite{}.  A 2D foreground weighted power spectrum is shown in Figure \ref{fig:weighted_2d}. The power spectra shown in Figure \ref{fig:1D_pspecs} show an excess of power in excess of the expected noise. This excess is notably similar between both calibration/foreground subtraction pipelines. The amount of power in the excess, as compared with the error bars, also depends rather dramatically on the range of $k$ bins included in the final averaging to the 1D.  These are discussed in more detail in section \ref{sec:power_spectrum_comparison}.

\subsection{Power Spectrum \#3: Empirical Covariance}
\label{sec:empirical_cov}

The \empirical{} power-spectrum estimation method computes both 2D and 1D power spectra using the quadratic estimator formalism. The method and its application to this data is described in more detail by \dilloncite{}.

The quadratic estimator method of \cite{Liu:2011p8763} treats foreground residuals in maps as a form of correlated noise and simultaneously downweights both noisy and foreground-dominated modes, keeping track of the extra variance they introduce into power spectrum estimates. This technique can be computationally demanding but using acceleration techniques described by \cite{Dillon:2013p10497}, has been applied to the previous MWA 32T results of \cite{Dillon:2014p9788} while a very similar technique, working on visibilities rather than maps but also using the data itself to estimate covariance, was used for the recent PAPER 64 results of \cite{2015ApJ...809...61A}.  \dilloncite{} build on these methods to mitigate errors introduced by imperfect mapmaking and instrument modeling through empirical covariance estimation, assuming all data covariance is sourced by foregrounds.

\empirical{} takes as input FHD calibrated images with foregrounds subtracted as well as possible, split into even and odd time-slices and averaged over many observations. The differences between the even and odd time-slices, which are assumed to be pure noise, are used to calibrate the system temperature in a noise model derived from observation time in $uv$ cells. In order to avoid directly propagating instrumental chromaticity into the foreground residual covariance models \citep{2015PhRvD..91b3002D}, \empirical{} uses the data itself to properly downweight residual foregrounds as seen by the instrument. It does this by estimating the frequency-frequency foreground residual covariance in annuli in $uvf$ space, assuming that different $uv$ cells independently sample the foreground residuals. This assumption, similar to that made by CHIPS, allows the combined foreground and noise covariance to be inverted directly and used to downweight the cubes when binning into 2D (and eventually 1D) band powers. As part of the quadratic estimator formalism, \empirical also calculates error covariances and window functions (i.e.\ horizontal error bars). The resulting power spectrum is shown in Figures \ref{fig:weighted_2d} and \ref{fig:1D_pspecs}.

\subsection{Benefits of Comparison}
\label{sec:benefits_of_comparison}

One benefit from having multiple pipelines is the freedom to investigate  different optimization axes.  The design of the \eppsilon{} power spectrum estimator emphasizes speed and relative simplicity, choices  motivated by the need to understand the effect, on the power spectrum, of processing decisions such as observation protocol, flagging, and calibration. Using \eppsilon{} we have discovered and corrected multiple systematic effects, primarily those of a spectral nature which were not obvious in imaging but quite apparent in the 2D power spectrum. With the ability to quickly form power spectra on different sets of data, \eppsilon{} has been an important tool for selecting sets of high quality data. 

In contrast, CHIPS starts from time-ordered data and in its calculations emphasizes a more full accounting of instrumental and residual foreground covariance. Not only does this higher resolution covariance calculation provide a more accurate accounting of the instrumental window function on the power spectrum, but it also allows for more precise weighting schemes based on knowledge of the statistical properties of the residual foregrounds. This is useful when making 1D power spectra where foreground-like modes can be downweighted in the average. 

Somewhere in the middle of these two is \empirical{} which, like \eppsilon{} uses image cubes and associated weighting variances but performs a more formal quadratic estimator in which additional covariances can be downweighted and the effects of the instrument window function be factored into the calculation of the power spectrum bins and error bars.  It also demonstrates the impact of inverse covariance weighting by forming a measure of covariance from the data. This measure encapsulates both the residual foregrounds modeled by CHIPS as well as any other residuals resulting from mis-calibration.

\section{Comparison Discussion}
\label{sec:results}

Inspecting a comparison of the images and power spectra reveals several common features. Images before and after foreground subtraction are shown in Figure \ref{fig:image_compare}, presented in the natural weighting used by the power spectrum estimators without application of any further cleaning.  Putting the same 3 hours of MWA data into each pipeline, we inspect output images before and after foreground subtraction. The pre foreground-subtracted (sometimes called the ``dirty'' image) have residuals at about the 15\% level; after foreground subtraction the differences are somewhat larger at 30\%. Residuals in the dirty maps are largest  around bright sources. This is most likely due to slight differences in the calculation of image plane weights which are dramatically emphasized by the broad psf from the natural weighting. As evidenced by the clean residual maps, the point source subtraction is well modeled when subtracted in the visibilities.  The foreground subtracted images (sometimes called ``residual'' images) show a much closer agreement both around the subtracted sources and in the large scale structure. Large scale structure is more difficult to distinguish. Inspection of the snapshot images before averaging in time and frequency revealed that the structure is consistent across both time and frequency, which suggests real Galactic emission rather than sidelobes or aliasing.

\subsection{Power Spectra}
\label{sec:power_spectrum_comparison}

We apply both \eppsilon{} and the unweighted version of CHIPS to both of our calibration and foreground subtraction pipes to produce a total of four different power spectra (Figures \ref{fig:pspec_compare} and \ref{fig:1d_kperp}).  Each power spectrum estimator has been developed to target the output from a ``primary'' calibration and foreground subtraction process---the diagonal panels of Figure \ref{fig:pspec_compare}---and have been highly optimized to that up-stream source of data.  The off-diagonal power spectra were created using auxiliary links which import the data and the metadata produced by the foreground subtraction step.  Since they are less highly optimized, lacking as they do the advantage of a close working relationship, these pathways represent an upper limit on the variance to be expected from small analysis differences but allow us to look for effects common to foreground subtraction or to power spectrum method.

Properties shared by all are the large amount of power at low $k_{\parallel}$ roughly at an amplitude of $10^{15}$ mK$^2$/(Mpc/h)$^3$. This emission is approximately flat over most of $k_{\perp}$ but rises steeply rise at low $k_\perp$. The amplitude agreement is particularly apparent in Figure \ref{fig:1d_kperp} where we plot a slice of the 2D power spectrum at $k_\parallel=0$ where most foreground power is expected to lie. A model of smooth galactic emission has not been subtracted which likely contributes to this steep rise. The ``wedge'' shaped linear dependance on baseline length in the 2D power spectra is due to the inherently chromatic response of a wide field instrument to smooth spectrum foregrounds; sources entering far from the phase center appear as bright pixels at higher $k_\parallel$ with sources on the horizon at the edge indicated by Figure \ref{fig:pspec_compare}'s  solid black line. The solid and dotted lines in the figure indicate the upper boundaries of power from sources at the horizon and at the beam half power point, respectively.  With the exception of some instrumental features foreground power is well isolated within this expected boundary. This emission is also visible in the image cubes as side-lobes extending from outside of the imaged area which move as function of frequency.  Observations recorded when the Galactic plane is near the horizon have a much larger wedge component and have been excluded from this analysis. See \cite{2015ApJ...804...14T}, \cite{2015ApJ...807L..28T} and \cite{Pober:2016ApJ...819....8P} for a detailed discussion of the foreground contributions to the power spectra in this data.

The two main instrumental systematics are horizontal striping due to missing or poorly calibrated data at the edges of regular coarse passbands and vertical striping due to spectral variation near uneven $uvf$ sampling. As described in section \ref{sec:observing}, the MWA reads out spectra which are divided into 1.28\,MHz wide ``coarse bands''. These bands have small known aliasing and gain instabilities in the edge channels and so during initial flagging we flag the edge-most 85kHz.  This regular gap in the spectra corresponds to a poor sampling in Fourier space at integer values of $\eta=781$ns or $k_\parallel=0.45$hMpc$^{-1}$.  In the 2D power spectrum this manifests as horizontal stripes of high power which are in fact sidelobes of the foreground wedge, and higher error bars reflecting lack of information about these modes.  These sidelobes can be minimized by working to improve the accuracy of the passband calibration and so have fewer flagged channels. They can also be downweighted by accounting for the covariance between modes as is done by \empirical{}. 

 A similar issue also arises from gaps in the $uv$ coverage. In places where coverage is not uniform --such as at longer baseline lengths where baseline density drops as $1/r^2$-- beam weights can vary dramatically as a function of frequency leading to a vertical striping effect. It is most prominent above $|u|>100\lambda$ or $k_\perp>0.1$hMpc$^{-1}$ which corresponds to when $uv$ sampling begins to drop below unity beyond the densest part of the MWA core. This effect can be ameliorated by using an accurate model for the beam and array positions to grid these samples according to the optimal mapmaking procedure, the approach taken by the FHD imager, or the CHIPS approach of calculating and inverting the full instrumental covariance.

The most noticeable difference between the different pipeline paths is in the power level in the window above the horizon and below the first coarse passband line (between 0.1 and 0.3 $k_\parallel$ and 0.01 and 0.05 $k_\perp$). FHD to \eppsilon{} displays a noise-like window in the 2D space, with a number of points dipping below zero while the other methods are noise like only at at much higher $k$s.  One commonality between all power spectra with this positive bias is a relatively higher amplitude of the coarse passband lines.

The relative amplitude of the vertical striping is probably the largest difference between the four power spectrum methods.   FHD-\eppsilon{} sees vertical striping largely consistent with noise, the other methods see the striping at varying levels with both CHIPS spectra showing the largest.  As discussed in detail by \cite{Morales:2012p8790} and shown in data by \citet{Pober:2016ApJ...819....8P}, this vertical striping is very sensitive to the accuracy of the weights used to average multiple samples together.  \eppsilon{} relies on the imager (FHD or RTS) to simulate the instrument and generate optimally weighted maps while CHIPS uses an internal instrument model to calculate covariance. FHD used the second generation \citet{Sutinjo:2015RaSc...50...52S} beam model which takes into account cross-coupling within a tile while the rest used an analytic short-dipole approximation.

\begin{figure}[htbp]
\begin{center}
\includegraphics[width=0.5\textwidth]{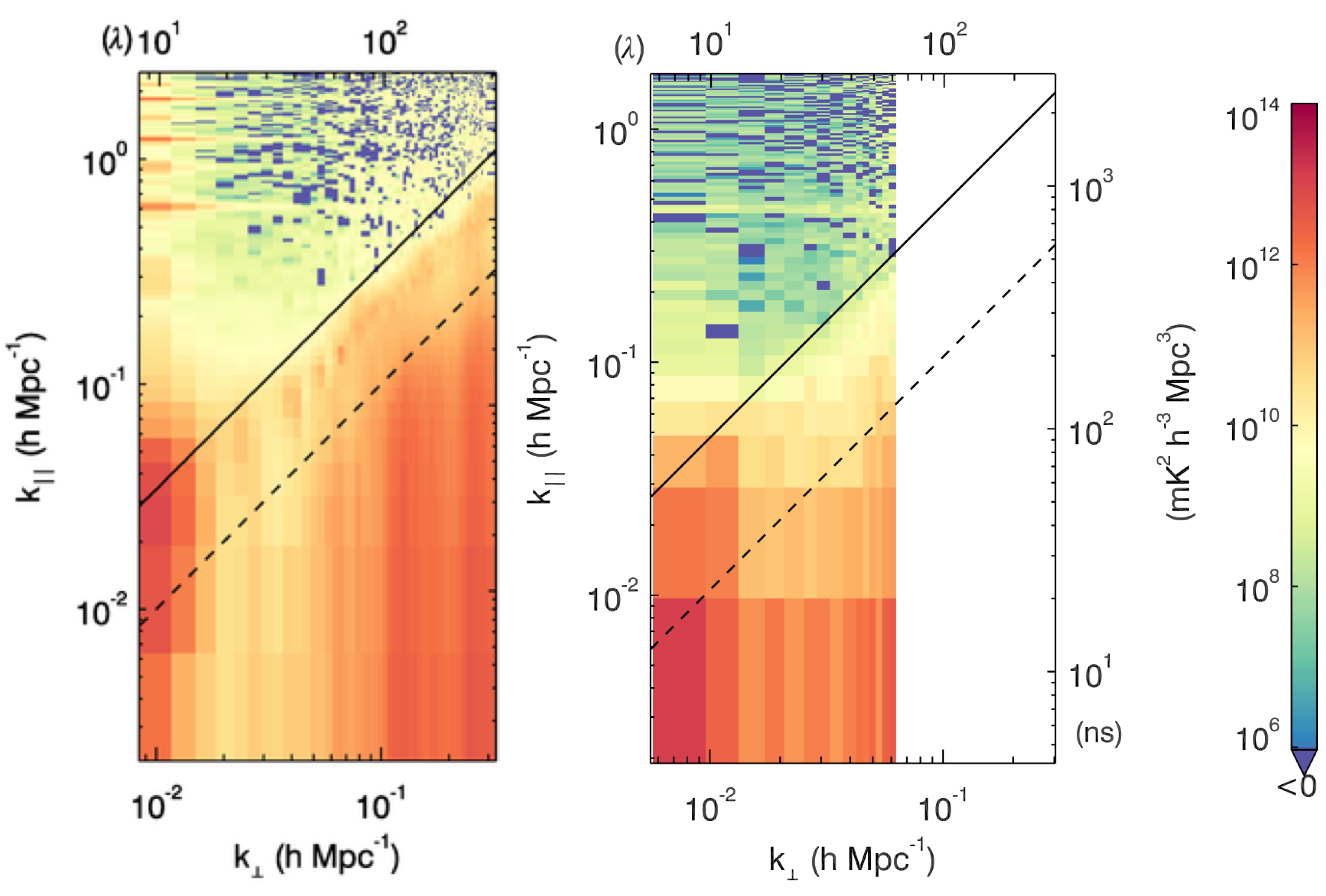}
\caption{\label{fig:weighted_2d} Choice of weighting scheme when applying inverse covariance weights can significantly reduce the effect of the foreground wedge on higher $k_\parallel$ modes.  On the left, a CHIPS power spectrum where inverse covariance includes a statistical model of confused foregrounds, on the right the \empirical{} estimator has weighted by an estimate of covariance formed from the data cube. }
\end{center}
\end{figure}

We also compare the results of CHIPS and \empirical{} using analogous foreground downweighting schemes. A quantitative comparison of these power spectra is difficult, since the quadratic estimator's downweighting scheme does not preserve foreground power. However, the results in Figure \ref{fig:weighted_2d} are largely similar, showing the familiar wedge structure and the brightest foreground contamination at low $k_\perp$ where galactic foregrounds dominate. \empirical{} excludes long baselines where coverage and sensitivity is poor and as such does not probe to the same range in $k_\perp$ as CHIPS. \empirical{} appears to more successfully remove foreground contamination near the wedge, which likely means that the foreground models employed in CHIPS have room for improvement. Likewise, \empirical{} can successfully remove the lines in constant $k_\parallel$ that arise from flagged channels due to the MWA's coarse band structure, but is still contaminated by the 90~m cable reflection at $k_\parallel \approx 0.45$ $h$Mpc$^{-1}$ (\dilloncite{}).

The final analysis step is to average into 1D power spectra along shells of constant $k$. These are shown in Figure \ref{fig:1D_pspecs} for three of the four analysis tracks\footnote{The RTS$\to$\eppsilon{} spectrum is excluded here because at the time of this analysis the RTS did not produce absolutely normalized image-plane uncertainties which are necessary to calculate 1D error bars.} shown in Figure \ref{fig:pspec_compare} with the addition of the \dilloncite{} points and a theoretical sensitivity curve calculated using the 21CMSENSE sensitivity code\footnote{\url{github.com/jpober/21cmsense/}} by \cite{Pober:2014p10390}.

The positive biases visible in the 2D power spectra are also apparent here. Only a few points are fully consistent with zero at 2$\sigma$; however, most are very close to the theoretical sensitivity curve and have errors matching those predicted for noise.
 The power spectra fall into two groups, those calculated from input image cubes (\eppsilon{} and \empirical{}) and those calculated directly from visibilities using CHIPS. The image-based points are somewhat deeper at low $k$, as noted in the 2D plots.  Points from CHIPS are biased more strongly at low $k$ but the slope is flatter and converges with the other pipelines at higher $k$.  

\subsection{CHIPS bias and the interpretation of error bars}
\begin{figure*}[htbp]
\begin{center}
\includegraphics[width=0.8\textwidth]{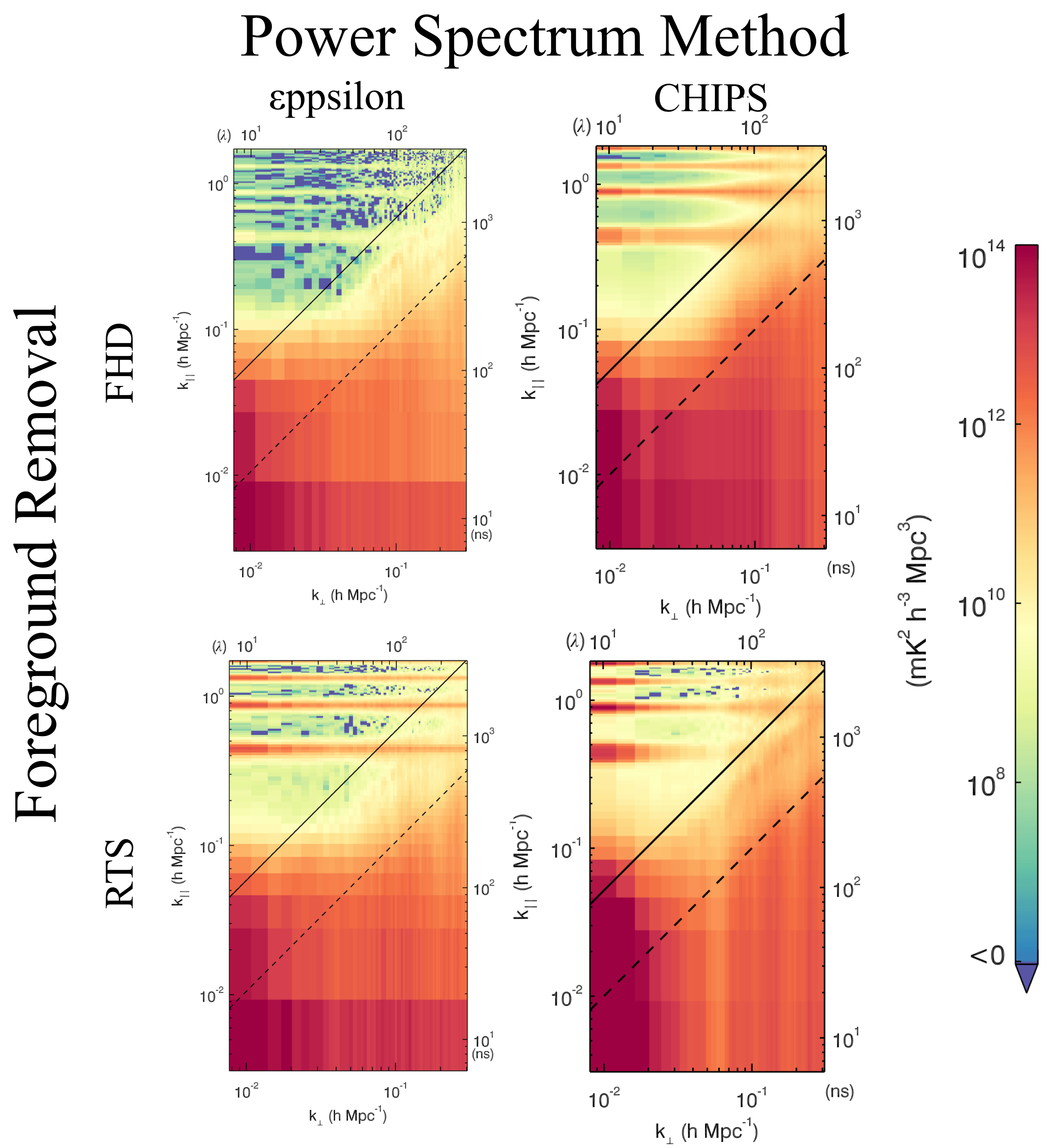}
\caption{Power spectra computed using two foreground subtraction methods and two power spectrum estimation methods on the data shown in Figure \ref{fig:image_compare}; the power spectrum has been computed in 3D spectral line cubes and then averaged cylindrically.  In the top row data have been calibrated and foreground subtracted using the Fast Holographic Deconvolution method, in the bottom row by the MWA Real Time System.  In the left column, power spectra have been estimated with \eppsilon{}, which emphasizes speed and full error propagation, in the right column, CHIPS corrects more correlation between $k$ modes.  All spectra display the now well-understood ``wedge''-shaped foreground residual and horizontal stripes caused by evenly spaced gaps in the instrument pass-band. Because all the power spectra are calculated by cross-multiplying independent data samples, measurement noise remains zero mean; negative regions are therefore indicative of noise-dominated regions. 
In these estimates no additional downweighting of foreground modes has been performed. See Fig   \ref{fig:weighted_2d} for an example of the effect of applying a more aggressive weighting scheme with \empirical{}.
\label{fig:pspec_compare}}
\end{center}
\end{figure*}

\begin{figure}[htbp]
\begin{center}
\includegraphics[width=0.5\textwidth]{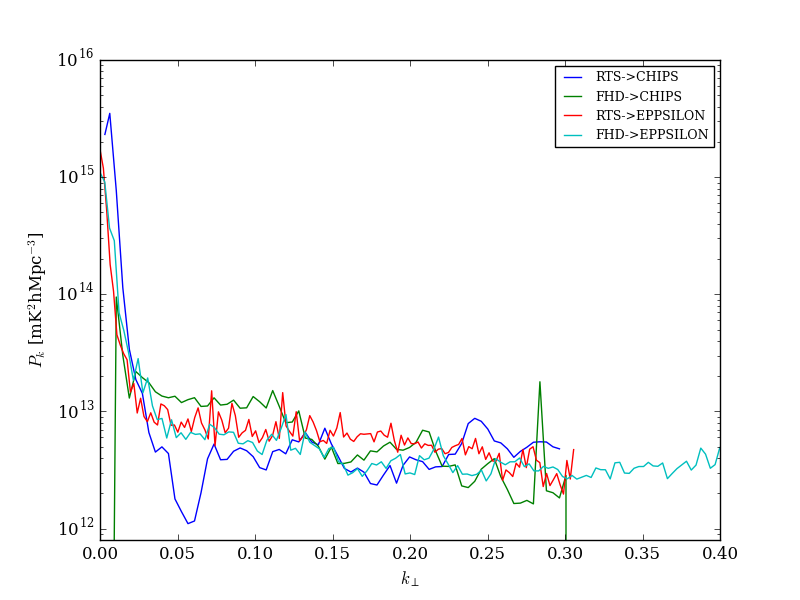}
\caption{Horizontal cut sampling the $k_\parallel = 0$ mode of the 2D power spectra shown in Figure \ref{fig:pspec_compare} indicating good agreement on flux scale and foreground power shape over most k modes. The foreground subtraction model only includes point sources. The steep rise is likely due to the bright, smooth galactic foreground emission visible in the residual images in Figure \ref{fig:image_compare} and power spectra by \cite{2015ApJ...807L..28T}. Power spectra produced by \empirical{} did not include the $k_\parallel=0$ mode.}
\label{fig:1d_kperp}
\end{center}
\end{figure}

\begin{figure*}[htbp]

\includegraphics[width=\textwidth]{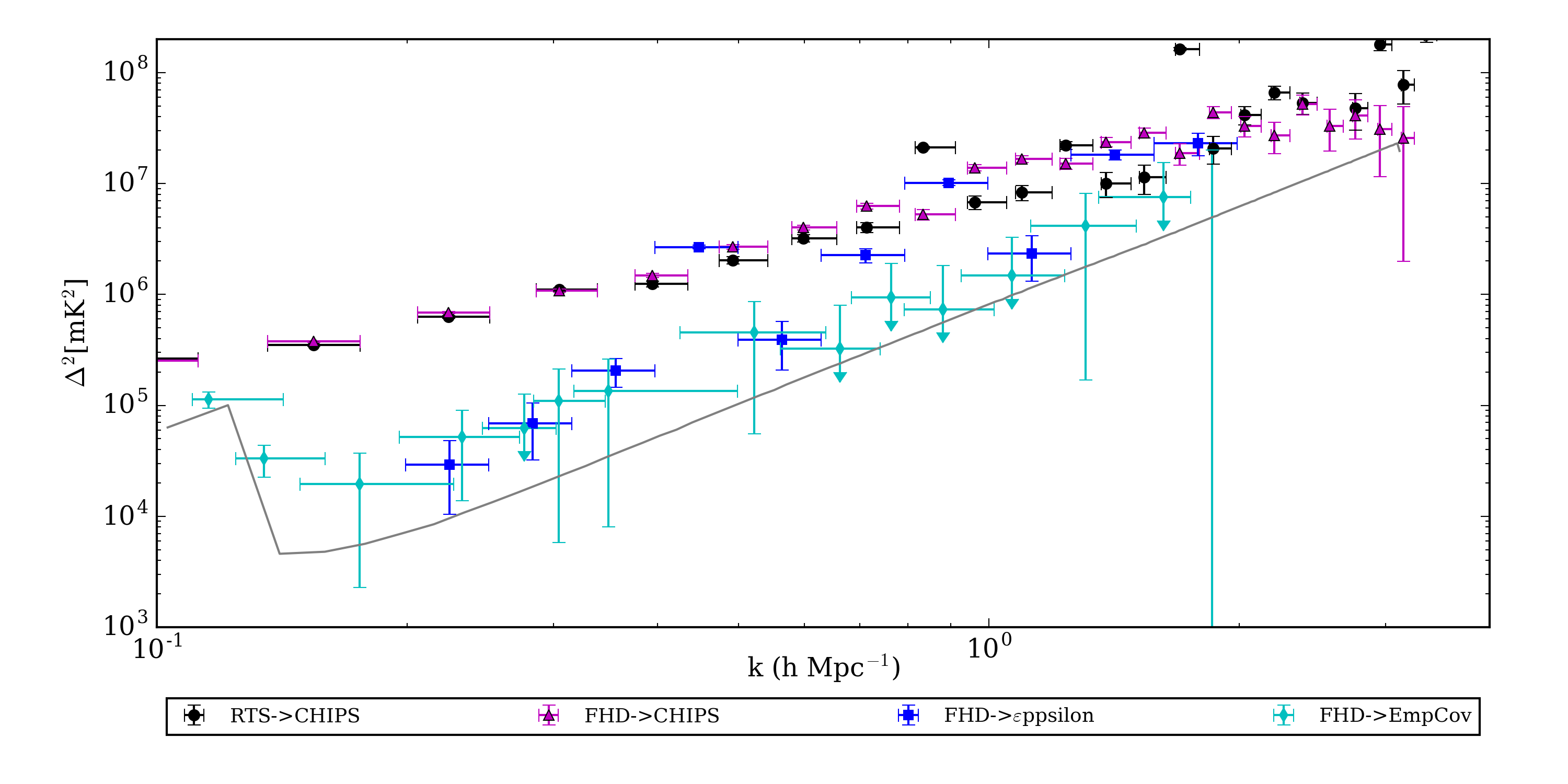}

\caption{Power spectra averaged along shells of constant $k$ with 2$\sigma$ errors. In three hours of data, four different methods demonstrate different kinds of limits on the power spectrum. Note that of the four pathways shown in Figure \ref{fig:pspec_compare}, only three are included here, but we have now included the \empirical{} power spectrum from \dilloncite{}.  \eppsilon{} power spectra of RTS outputs are not shown because, in the version under test here, RTS did not natively produce image-plane error bars, which are required to correctly average from 2D to 1D. Many of the features visible in the 2D plots are also visible here: the excess in the CHIPS spectra is clearly visible as is a smaller excess in the \eppsilon{} spectrum.  The black line indicates 2$\sigma$ bounds for points dominated by noise.  Power levels for typical theoretical models are typically in the 5 to 10 mK$^2$ range across these $k$ modes.
\label{fig:1D_pspecs}}

\end{figure*}

Part of the CHIPS bias is due to the calculation of weightings. Default CHIPS analysis uses a statistical model of confused foregrounds to downweight biased modes, particularly those correlated with the wedge power. For this reason it is desirable to include the wedge modes in that 1D average. However it significantly changes the interpretation of error bars; points in which a significant amount of power have been downweighted will have error bars much larger than thermal. In the interest of comparing with the other methods, the power spectra in Figures \ref{fig:pspec_compare}, \ref{fig:1d_kperp} and \ref{fig:1D_pspecs} have been calculated using points only lying outside the wedge horizon. This limits the amount of wedge-to-window covariance CHIPS can remove and contributes to the larger bias. 

\begin{figure*}[htbp]
\begin{center}
\includegraphics[width=\textwidth]{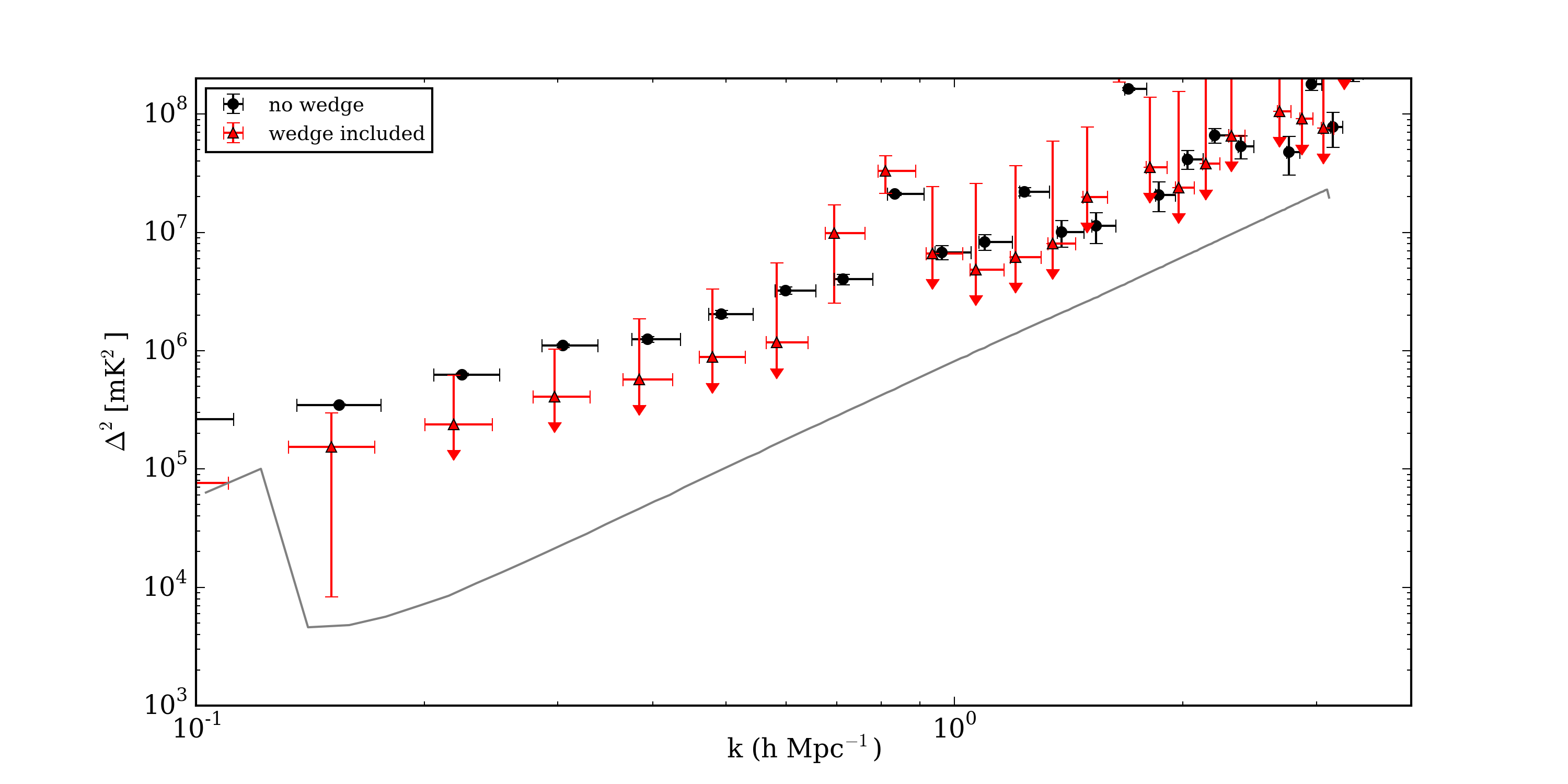}
\caption{An example of the dramatic impact that weighting and covariance minimization has on the interpretation of error bars.  Here we compare the RTS$\to$CHIPS power spectrum from Figure \ref{fig:1D_pspecs} with that given by \chipscite{}. The latter was made with the same data but only 1/3 of the 30MHz band, and so slightly higher error bars.  In both, CHIPS downweights by a model of foreground covariance formed by propagating a statistical model of confused sources. The only difference is that black excludes the wedge but red does not. When the wedge is included the modeled foregrounds in those $k$-space voxels dominate the covariance weights. Applying these weights essentially moves the foreground bias into the error bars and asserts that, given our best model of foregrounds, the power spectra are completely consistent with noise and foregrounds and do not provide evidence for a significant cosmological 21 cm signal.}
\label{fig:CHIPS_compare}
\end{center}
\end{figure*}

Including the full wedge in the CHIPS covariance calculation offers foreground suppression, but also introduces a foreground component into the error bars. Compare in Figure \ref{fig:CHIPS_compare} the RTS$\to$CHIPS power spectrum in Figure \ref{fig:1D_pspecs} with that given by \chipscite{} which used the same data shown here, though only 1/3 of the 30MHz band.  In both, CHIPS has downweighted by a model of foreground covariance formed by propagating a statistical model of confused sources. The only difference is that black excludes the wedge but red does not. When the wedge is included the modeled foregrounds in those voxels dominate the covariance weights. Applying these weights essentially moves the foreground bias into the error bars and asserts that, given our best model of foregrounds, the power spectra are completely consistent with noise and  foregrounds.

The power spectra in Figure \ref{fig:1D_pspecs} show the range of results possible given the same input data. Though they do not all agree, they do paint a consistent picture.  Differences partly come from the definition of error bars but also indicate the relative difficulty of methods. Methods which rely on an imager seem to perform somewhat better. This is perhaps unsurprising. CHIPS computes the instrumental correlation matrix in visibility space using beam, bandpass, etc. As the CHIPS analysis exists entirely in the visibility space, errors in modeling the instrument are perhaps more difficult to detect than they are in the image space.  However, we do not suggest that visibility-based calculations like CHIPS are doomed to failure; rather the opposite. The instrument models will continue to improve, and this improvement will be easily validated by comparison with the other pipelines.

\section{Lessons from Comparing Independent Pipelines}
\label{sec:lessons}
          A data analysis pipeline is necessarily built on a complex software framework which is only approximately described in prose; it is therefore both difficult to perfectly replicate and susceptible to human error.  Comparison between independently developed analysis paths, each with their own strengths and limitations, is essential to placing believable constraints on the Epoch of Reionization. The ongoing comparison between independent MWA pipelines has revealed a number of issues both systematic (related to our understanding of the instrument or foregrounds) and algorithmic (optimizing our use of this knowledge) which we will briefly mention here.
      
\begin{itemize}

\item \emph{ Systematic example: cable reflections}

As discussed above, one significant difference between the two pipelines is the number of free parameters fit in the calibration step, particularly in the spectral dimension.  Both calibration pipelines begin by calibrating each channel and then averaging over a number of axes.  The RTS fits a low order polynomial, piecewise, to each of the 24 1.28MHz sub-band solutions, while FHD fits a similar order polynomial to the entire  band's calibration solution.  Inspection of power spectra calibrated using the FHD scheme revealed previously unknown spectral features corresponding to reflections on the analog cables at the -20dB level (~1.5\%). FHD calibration now includes a fit for these reflections and the feature is substantially reduced. These features are fully covered by the RTS fit (which uses of order 10 times as many free parameters as FHD). 

\item \emph{Calibration example: number of sources}

In early comparisons between RTS and FHD images one immediately apparent difference was the somewhat lower dynamic range of the RTS images.  This was traced to the largest (at the time) difference between the two approaches; RTS used the more traditional radio astronomy practice of calibrating to a pointing on a bright source at the beginning of each night and then transferring the calibration to the rest of the observations, whereas FHD was calibrating against the foreground model using the cataloged sources within the field of view (a few thousand). This dramatically highlighted the breakdown of approaches designed for traditional dish telescopes with a narrow field of view.  The MWA field of view is so wide, that even the calibration pointing included many sources of brightness comparable to the calibrator. These sources were not included in the calibration model and thus limited the accuracy of the calibration. Also, owing to the phased array beam steering, the primary beam for the calibrator pointing is very different from the beams used for the primary reionization observations, particularly in polarization response.  So, though the instrument itself is highly stable in time over many hours, calibrations must be carefully matched up with the observing parameters or experience a dramatic loss of imaging dynamic range, both spatially and spectrally. The addition of ``in field'' calibration, where the foreground subtraction model is also the calibration model, significantly improved the RTS images and brought the two imagers into substantial agreement.

\item \emph{Algorithmic example: full forward modeling for absolute calibration and signal loss}

During the comparison process, one way in which all pipeline results differed from each other is in the overall amplitude of the power spectrum scale. Flux calibration, weightings, Fourier conventions and signal loss must all be well understood for good agreement to be reached.  Signal loss, in particular, must be examined closely. Unintentional or unavoidable down-weighting or subtraction of reionization signal could occur at multiple stages such as bandpass calibration, $uvf$ gridding, or inverse covariance weighting. These effects are best calibrated via forward modeling of simulated sky inputs.  For example, detailed simulations of reionization signals through FHD and \eppsilon{} found that in areas of dense $uv$ sampling, simulated power spectra experienced a 50\% reduction of detected power (\eppsiloncite{}). The act of gridding complex visibilities into the uv-plane with a convolving kernel does not conserve the overall normalization of the power spectrum.  This effect has been confirmed with simulations and results in a factor of 2 correction in the power spectrum; see \eppsiloncite{} in prep. for a more detailed explanation.

These simulated reionization data sets have been calibrated internally by comparing outputs at every step of the imaging and power spectrum process, and so are well understood at a detailed level, and suitable for use as calibration standards for new pipelines.

\item \emph{Algorithmic example: w-planes in power spectrum calculation}

Many of the differences found between power spectra during the comparison were traced to the post-foreground-subtraction steps, particularly the implementation of new imaging and power spectrum estimation codes.  One example was an anomalous loss of power in CHIPS power spectra which particularly affected longer baselines.  CHIPS grids in a coordinate space defined by the baseline vector $\vec{b}$ and spectral mode $\eta$ and then uses an instrument model to diagonalize and sum in this sparse power spectrum space.  Unlike FHD which uses snapshots to avoid directly handling the third or `w' term of the baseline vector, CHIPS accumulates the entire observation into a full uvw$\eta$ hyper-cube.  The number and size of the voxels in this space, particularly in the w direction is a somewhat free parameter and relates to the precision of the instrument model, the amount of time included and other factors.  Subsequent, more detailed foreground simulations suggested a factor of 4 w resolution increase which eliminated the signal loss and dramatically improved agreement.

\vfill
\break

\item \emph{Interchange standards}

Finally, in the interest of transparency, we offer a somewhat prosaic but perhaps vital lesson regarding  nomenclature. For fixed dipole arrays there are (at least) two popular and mutually exclusive traditions. Tradition A: In keeping with the customary abscissa of latitude longitude plots, the east-west oriented dipole is labeled X.  Tradition B: Astronomically, the X polarization is measured as the amplitude of a dipole aligned with lines of constant Right Ascension; which for a source at zenith maps to north-south. We humbly suggest that those pursuing a cross comparison effort select one standard at the outset.  

\end{itemize}

  We must stress that without the ability to compare between independent pipelines, most of these effects would have gone un-detected or mis-diagnosed as algorithmic deficiencies and have persisted into the final result or motivated additional fitting parameters resulting in higher signal loss as well as a vague disquiet. In addition to pipeline redundancy, forward modeling can provide some important checks, for example the absolute calibration of FHD and \eppsilon{} described in \eppsiloncite{}, however the result is only as good as the model itself.
  
\section{Conclusions}
\label{sec:conclusion}
In this overview paper we have provided a top level view of foreground subtraction and power spectrum estimation methods of the MWA Epoch of Reionization project, described more completely in companion papers \eppsiloncite{}, \chipscite{}, and \dilloncite{}.  In this comparison we see that both foreground subtraction methods are able to reliably remove similar amounts of power.  Differences between the images are smaller than the remaining residual foregrounds by a factor of 3.2, suggesting an overall $\sim$30\% error on the aggregate calibration, foreground subtraction and imaging between the two pipelines.  The power spectra of these foreground subtracted outputs agree on the scale and distribution of power, though with some differences in the leakage of power into the window. These differences are partly due to definition of error bars and whether they include just noise or also foreground terms.  

Including foregrounds in the error calculation is a key exercise because it lets us answer more nuanced questions. Rather than simply: is the data inconsistent with a 21cm detection in the presence of noise? With CHIPS we can ask: Is the data inconsistent with a 21cm detection in the presence of noise and an a-priori foreground model? With \empirical{} we can ask: is the data inconsistent with a 21 cm detection in the presence of noise and a foreground model fit to the data? These are all good questions.

21cm cosmology experiments have very wide fields of view, dense samplings, drift scanning observing and the reionization science levies a requirement for very high---at least 10,000:1---spectral dynamic range. All of this has necessitated the development of new algorithms for calibration and imaging, as well as the surrounding scaffolding to process thousands of hours of data to achieve this precision.  This paper is the first step towards validating these pipelines and providing robust repeatable results.

\acknowledgments

We thank our referee for a comprehensive review which materially improved the paper. This work was supported	 by the U. S. National Science Foundation (NSF) through award AST--1109257. DCJ is supported by an NSF Astronomy and Astrophysics Postdoctoral Fellowship under award AST--1401708. CMT is supported by an Australian Research Council DECRA Award, DE140100316. This scientific work makes use of the Murchison Radio-astronomy Observatory, operated by CSIRO. We acknowledge the Wajarri Yamatji people as the traditional owners of the Observatory site. Support for the MWA comes from the U.S. National Science Foundation (grants  AST-1410484, AST-0821321, AST-0457585, PHY-0835713, CAREER-0847753, and AST-0908884), the Australian Research Council (LIEF grants LE0775621 and LE0882938), the U.S. Air Force Office of Scientific Research (grant FA9550-0510247), MIT School of Science, the Marble Astrophysics Fund, and the Centre for All-sky Astrophysics (an Australian Research Council Centre of Excellence funded by grant CE110001020). Support is also provided by the Smithsonian Astrophysical Observatory, the MIT School of Science, the Raman Research Institute, the Australian National University, and the Victoria University of Wellington (via grant MED-E1799 from the New Zealand Ministry of Economic Development and an IBM Shared University Research Grant). The Australian Federal government provides additional support via the Commonwealth Scientific and Industrial Research Organisation (CSIRO), National Collaborative Research Infrastructure Strategy, Education Investment Fund, and the Australia India Strategic Research Fund, and Astronomy Australia Limited, under contract to Curtin University. We acknowledge the iVEC Petabyte Data Store, the Initiative in Innovative Computing and the CUDA Center for Excellence sponsored by NVIDIA at Harvard University, and the International Centre for Radio Astronomy Research (ICRAR), a Joint Venture of Curtin University and The University of Western Australia, funded by the Western Australian State government.

\bibliography{library}

\end{document}

%% file: mwa_eor_collab.tex
\author{
Daniel~C.~Jacobs\altaffilmark{1}$^,$\altaffilmark{2},
B.~J.~Hazelton\altaffilmark{3}$^,$\altaffilmark{4},
C.~M.~Trott\altaffilmark{5}$^,$\altaffilmark{6},
Joshua~S.~Dillon\altaffilmark{7},
B.~Pindor\altaffilmark{5}$^,$\altaffilmark{8},
I.~S.~Sullivan\altaffilmark{3},
J.~C.~Pober\altaffilmark{9},
N.~Barry\altaffilmark{3},
A.~P.~Beardsley\altaffilmark{1}$^,$\altaffilmark{3},
G.~Bernardi\altaffilmark{10}$^,$\altaffilmark{11}$^,$\altaffilmark{12},
Judd~D.~Bowman\altaffilmark{1},
F.~Briggs\altaffilmark{5}$^,$\altaffilmark{13},
R.~J.~Cappallo\altaffilmark{14},
P.~Carroll\altaffilmark{3},
B.~E.~Corey\altaffilmark{14},
A.~de~Oliveira-Costa\altaffilmark{7},
D.~Emrich\altaffilmark{6},
A.~Ewall-Wice\altaffilmark{7},
L.~Feng\altaffilmark{7},
B.~M.~Gaensler\altaffilmark{15}$^,$\altaffilmark{5}$^,$\altaffilmark{16},
R.~Goeke\altaffilmark{7},
L.~J.~Greenhill\altaffilmark{12},
J.~N.~Hewitt\altaffilmark{7},
N.~Hurley-Walker\altaffilmark{6},
M.~Johnston-Hollitt\altaffilmark{17},
D.~L.~Kaplan\altaffilmark{18}$^,$\altaffilmark{3},
J.~C.~Kasper\altaffilmark{19}$^,$\altaffilmark{12},
HS Kim\altaffilmark{5}$^,$\altaffilmark{8},
E.~Kratzenberg\altaffilmark{14},
E.~Lenc\altaffilmark{5}$^,$\altaffilmark{16},
J.~Line\altaffilmark{5}$^,$\altaffilmark{8},
A.~Loeb\altaffilmark{12},
C.~J.~Lonsdale\altaffilmark{14},
M.~J.~Lynch\altaffilmark{6},
B.~McKinley\altaffilmark{5}$^,$\altaffilmark{8},
S.~R.~McWhirter\altaffilmark{14},
D.~A.~Mitchell\altaffilmark{20}$^,$\altaffilmark{5},
M.~F.~Morales\altaffilmark{3},
E.~Morgan\altaffilmark{7},
A.~R.~Neben\altaffilmark{7},
N.~Thyagarajan\altaffilmark{1},
D.~Oberoi\altaffilmark{21},
A.~R.~Offringa\altaffilmark{5}$^,$\altaffilmark{22},
S.~M.~Ord\altaffilmark{5}$^,$\altaffilmark{6},
S. Paul\altaffilmark{23},
T.~Prabu\altaffilmark{23},
P.~Procopio\altaffilmark{5}$^,$\altaffilmark{8},
J.~Riding\altaffilmark{5}$^,$\altaffilmark{8},
A.~E.~E.~Rogers\altaffilmark{14},
A.~Roshi\altaffilmark{24},
N.~Udaya~Shankar\altaffilmark{23},
Shiv~K.~Sethi\altaffilmark{23},
K.~S.~Srivani\altaffilmark{23},
R.~Subrahmanyan\altaffilmark{5}$^,$\altaffilmark{23},
M.~Tegmark\altaffilmark{7},
S.~J.~Tingay\altaffilmark{5}$^,$\altaffilmark{6},
M.~Waterson\altaffilmark{13}$^,$\altaffilmark{6},
R.~B.~Wayth\altaffilmark{5}$^,$\altaffilmark{6},
R.~L.~Webster\altaffilmark{5}$^,$\altaffilmark{8},
A.~R.~Whitney\altaffilmark{14},
A.~Williams\altaffilmark{6},
C.~L.~Williams\altaffilmark{7},
C.~Wu\altaffilmark{25}$^,$\altaffilmark{3},
J.~S.~B.~Wyithe\altaffilmark{5}$^,$\altaffilmark{8}
}
\altaffiltext{1}{Arizona State University, School of Earth and Space Exploration, Tempe, AZ 85287, USA}
\altaffiltext{2}{e-mail: daniel.c.jacobs@asu.edu}
\altaffiltext{3}{University of Washington, Department of Physics, Seattle, WA 98195, USA}
\altaffiltext{4}{University of Washington, eScience Institute, Seattle, WA 98195, USA}
\altaffiltext{5}{ARC Centre of Excellence for All-sky Astrophysics (CAASTRO)}
\altaffiltext{6}{International Centre for Radio Astronomy Research, Curtin University, Perth, WA 6845, Australia}
\altaffiltext{7}{MIT Kavli Institute for Astrophysics and Space Research, Cambridge, MA 02139, USA}
\altaffiltext{8}{The University of Melbourne, School of Physics, Parkville, VIC 3010, Australia}
\altaffiltext{9}{Brown University, Department of Physics, Providence, RI 02912, USA}
\altaffiltext{10}{Department of Physics and Electronics, Rhodes University, Grahamstown 6140, South Africa}
\altaffiltext{11}{Square Kilometre Array South Africa (SKA SA), Park Road, Pinelands 7405, South Africa}
\altaffiltext{12}{Harvard-Smithsonian Center for Astrophysics, Cambridge, MA 02138, USA}
\altaffiltext{13}{Australian National University, Research School of Astronomy and Astrophysics, Canberra, ACT 2611, Australia}
\altaffiltext{14}{MIT Haystack Observatory, Westford, MA 01886, USA}
\altaffiltext{15}{Dunlap Institute for Astronomy and Astrophysics, University of Toronto, ON M5S 3H4, Canada}
\altaffiltext{16}{The University of Sydney, Sydney Institute for Astronomy, School of Physics, NSW 2006, Australia}
\altaffiltext{17}{Victoria University of Wellington, School of Chemical \& Physical Sciences, Wellington 6140, New Zealand}
\altaffiltext{18}{University of Wisconsin--Milwaukee, Department of Physics, Milwaukee, WI 53201, USA}
\altaffiltext{19}{University of Michigan, Department of Atmospheric, Oceanic and Space Sciences, Ann Arbor, MI 48109, USA}
\altaffiltext{20}{CSIRO Astronomy and Space Science (CASS), PO Box 76, Epping, NSW 1710, Australia}
\altaffiltext{21}{National Centre for Radio Astrophysics, Tata Institute for Fundamental Research, Pune 411007, India}
\altaffiltext{22}{Netherlands Institute for Radio Astronomy (ASTRON), PO Box 2, 7990 AA Dwingeloo, The Netherlands}
\altaffiltext{23}{Raman Research Institute, Bangalore 560080, India}
\altaffiltext{24}{National Radio Astronomy Observatory, Charlottesville and Greenbank, USA}
\altaffiltext{25}{International Centre for Radio Astronomy Research, University of Western Australia, Crawley, WA 6009, Australia}

%% file: MWA_Pipelines.bbl
\begin{thebibliography}{75}
\expandafter\ifx\csname natexlab\endcsname\relax\def\natexlab#1{#1}\fi

\bibitem[{{Ali} {et~al.}(2015){Ali}, {Parsons}, {Zheng}, {Pober}, {Liu},
  {Aguirre}, {Bradley}, {Bernardi}, {Carilli}, {Cheng}, {DeBoer}, {Dexter},
  {Grobbelaar}, {Horrell}, {Jacobs}, {Klima}, {MacMahon}, {Maree}, {Moore},
  {Razavi}, {Stefan}, {Walbrugh}, \& {Walker}}]{2015ApJ...809...61A}
{Ali}, Z.~S. {et~al.} 2015, \apj, 809, 61

\bibitem[{{Barry} {et~al.}(2016){Barry}, {Hazelton}, {Sullivan}, {Morales}, \&
  {Pober}}]{Barry:2016a}
{Barry}, N., {Hazelton}, B., {Sullivan}, I., {Morales}, M.~F., \& {Pober},
  J.~C. 2016, ArXiv e-prints

\bibitem[{Beardsley {et~al.}(2013)Beardsley, Hazelton, Morales, Arcus, Barnes,
  Bernardi, Bowman, Briggs, Bunton, Cappallo, Corey, Deshpande, Desouza,
  Emrich, Gaensler, Goeke, Greenhill, Herne, Hewitt, Johnston-Hollitt, Kaplan,
  Kasper, Kincaid, Koenig, Kratzenberg, Lonsdale, Lynch, Mcwhirter, Mitchell,
  Morgan, Oberoi, Ord, Pathikulangara, Prabu, Remillard, Rogers, Roshi, Salah,
  Sault, Udaya, Srivani, Stevens, Subrahmanyan, Tingay, Wayth, Waterson,
  Webster, Whitney, Williams, Williams, \& Wyithe}]{Beardsley:2013p9952}
Beardsley, A. {et~al.} 2013, Monthly Notices of the Royal Astronomical Society,
  429, L5

\bibitem[{Bernardi {et~al.}(2011)Bernardi, Mitchell, Ord, Greenhill, Pindor,
  Wayth, \& Wyithe}]{Bernardi:2011p9205}
Bernardi, G., Mitchell, D.~A., Ord, S.~M., Greenhill, L.~J., Pindor, B., Wayth,
  R.~B., \& Wyithe, J. S.~B. 2011, Monthly Notices of the Royal Astronomical
  Society, 413, 411

\bibitem[{{Bhatnagar} {et~al.}(2013){Bhatnagar}, {Rau}, \&
  {Golap}}]{Bhatnagar..2013ApJ}
{Bhatnagar}, S., {Rau}, U., \& {Golap}, K. 2013, \apj, 770, 91

\bibitem[{Bowman {et~al.}(2013)Bowman, Cairns, Kaplan, Murphy, Oberoi,
  Staveley-Smith, Arcus, Barnes, Bernardi, Briggs, Brown, Bunton, Burgasser,
  Cappallo, Chatterjee, Corey, Coster, Deshpande, deSouza, Emrich, Erickson,
  Goeke, Gaensler, Greenhill, Harvey-Smith, Hazelton, Herne, Hewitt,
  Johnston-Hollitt, Kasper, Kincaid, Koenig, Kratzenberg, Lonsdale, Lynch,
  Matthews, Mcwhirter, Mitchell, Morales, Morgan, Ord, Pathikulangara, Prabu,
  Remillard, Robishaw, Rogers, Roshi, Salah, Sault, Shankar, Srivani, Stevens,
  Subrahmanyan, Tingay, Wayth, Waterson, Webster, Whitney, Williams, Williams,
  \& Wyithe}]{Bowman:2013p9950}
Bowman, J. {et~al.} 2013, Publications of the Astronomical Society of
  Australia, 30, 31

\bibitem[{Bowman {et~al.}(2009)Bowman, Morales, \& Hewitt}]{Bowman:2009p7816}
Bowman, J., Morales, M., \& Hewitt, J. 2009, The Astrophysical Journal, 695,
  183

\bibitem[{Chapman {et~al.}(2012)Chapman, Abdalla, Harker, Jeli{\'c},
  Labropoulos, Zaroubi, Brentjens, Bruyn, \& Koopmans}]{Chapman:7p8505}
Chapman, E. {et~al.} 2012, Monthly Notices of the Royal Astronomical Society,
  423, 2518

\bibitem[{Chapman {et~al.}(2013)Chapman, Abdalla, Bobin, Starck, Harker,
  Jeli{\'c}, Labropoulos, Zaroubi, Brentjens, de~Bruyn, \&
  Koopmans}]{Chapman:2013p10379}
---. 2013, Monthly Notices of the Royal Astronomical Society, 429, 165

\bibitem[{{Chapman} {et~al.}(2012){Chapman}, {Abdalla}, {Harker}, {Jeli{\'c}},
  {Labropoulos}, {Zaroubi}, {Brentjens}, {de Bruyn}, \&
  {Koopmans}}]{2012MNRAS.Chapman.423.2518C}
{Chapman}, E. {et~al.} 2012, \mnras, 423, 2518

\bibitem[{Condon {et~al.}(1998)Condon, Cotton, Greisen, Yin, Perley, Taylor, \&
  Broderick}]{Condon:1998p7986}
Condon, J.~J., Cotton, W.~D., Greisen, E.~W., Yin, Q.~F., Perley, R.~A.,
  Taylor, G.~B., \& Broderick, J.~J. 1998, The Astronomical Journal, 115, 1693

\bibitem[{{Cornwell} {et~al.}(2012){Cornwell}, {Voronkov}, \&
  {Humphreys}}]{2012SPIE.8500E..0LC}
{Cornwell}, T.~J., {Voronkov}, M.~A., \& {Humphreys}, B. 2012, in \procspie,
  Vol. 8500, Image Reconstruction from Incomplete Data VII, 85000L

\bibitem[{Datta {et~al.}(2010)Datta, Bowman, \& Carilli}]{Datta:2010p8781}
Datta, A., Bowman, J.~D., \& Carilli, C.~L. 2010, The Astrophysical Journal,
  724, 526

\bibitem[{de~Oliveira-Costa {et~al.}(2008)de~Oliveira-Costa, Tegmark, Gaensler,
  Jonas, Landecker, \& Reich}]{deOliveiraCosta:2008p2242}
de~Oliveira-Costa, A., Tegmark, M., Gaensler, B.~M., Jonas, J., Landecker,
  T.~L., \& Reich, P. 2008, Monthly Notices of the Royal Astronomical Society,
  388, 247, (c) Journal compilation {\copyright} 2008 RAS

\bibitem[{Dillon {et~al.}(2013)Dillon, Liu, \& Tegmark}]{Dillon:2013p10497}
Dillon, J., Liu, A., \& Tegmark, M. 2013, Physical Review D, 87, 43005

\bibitem[{Dillon {et~al.}(2014)Dillon, Liu, Williams, Hewitt, Tegmark, Morgan,
  Levine, Morales, Tingay, Bernardi, Bowman, Briggs, Cappallo, Emrich,
  Mitchell, Oberoi, Prabu, Wayth, \& Webster}]{Dillon:2014p9788}
Dillon, J. {et~al.} 2014, Physical Review D, 89, 23002

\bibitem[{Dillon {et~al.}(2015)Dillon, Neben, Hewitt, Tegmark, Barry,
  Beardsley, Bowman, Briggs, Carroll, de~Oliveira-Costa, Ewall-Wice, Feng,
  Greenhill, Hazelton, Hernquist, Hurley-Walker, Jacobs, Kim, Kittiwisit, Lenc,
  Line, Loeb, McKinley, Mitchell, Morales, Offringa, Paul, Pindor, Pober,
  Procopio, Riding, Sethi, Shankar, Subrahmanyan, Sullivan, Thyagarajan,
  Tingay, Trott, Wayth, Webster, Wyithe, Bernardi, Cappallo, Deshpande,
  Johnston-Hollitt, Kaplan, Lonsdale, McWhirter, Morgan, Oberoi, Ord, Prabu,
  Srivani, Williams, \& Williams}]{PhysRevD.91.123011}
Dillon, J.~S. {et~al.} 2015, Phys. Rev. D, 91, 123011

\bibitem[{{Dillon} {et~al.}(2015){Dillon}, {Tegmark}, {Liu}, {Ewall-Wice},
  {Hewitt}, {Morales}, {Neben}, {Parsons}, \& {Zheng}}]{2015PhRvD..91b3002D}
{Dillon}, J.~S. {et~al.} 2015, \prd, 91, 023002

\bibitem[{Furlanetto {et~al.}(2006)Furlanetto, Oh, \&
  Briggs}]{Furlanetto:2006p2267}
Furlanetto, S.~R., Oh, S.~P., \& Briggs, F.~H. 2006, Physics Reports, 433, 181,
  elsevier B.V.

\bibitem[{G{\'o}rski {et~al.}(2005)G{\'o}rski, Hivon, Banday, Wandelt, Hansen,
  Reinecke, \& Bartelmann}]{Gorski:2005p7667}
G{\'o}rski, K., Hivon, E., Banday, A., Wandelt, B., Hansen, F., Reinecke, M.,
  \& Bartelmann, M. 2005, The Astrophysical Journal, 622, 759

\bibitem[{{Harker} {et~al.}(2009){Harker}, {Zaroubi}, {Bernardi}, {Brentjens},
  {de Bruyn}, {Ciardi}, {Jeli{\'c}}, {Koopmans}, {Labropoulos}, {Mellema},
  {Offringa}, {Pandey}, {Schaye}, {Thomas}, \&
  {Yatawatta}}]{Harker:2009MNRAS.397.1138H}
{Harker}, G. {et~al.} 2009, \mnras, 397, 1138

\bibitem[{{Heald} {et~al.}(2015){Heald}, {Pizzo}, {Orr{\'u}}, {Breton},
  {Carbone}, {Ferrari}, {Hardcastle}, {Jurusik}, {Macario}, {Mulcahy},
  {Rafferty}, {Asgekar}, {Brentjens}, {Fallows}, {Frieswijk}, {Toribio},
  {Adebahr}, {Arts}, {Bell}, {Bonafede}, {Bray}, {Broderick}, {Cantwell},
  {Carroll}, {Cendes}, {Clarke}, {Croston}, {Daiboo}, {de Gasperin}, {Gregson},
  {Harwood}, {Hassall}, {Heesen}, {Horneffer}, {van der Horst}, {Iacobelli},
  {Jeli{\'c}}, {Jones}, {Kant}, {Kokotanekov}, {Martin}, {McKean}, {Morabito},
  {Nikiel-Wroczy{\'n}ski}, {Offringa}, {Pandey}, {Pandey-Pommier}, {Pietka},
  {Pratley}, {Riseley}, {Rowlinson}, {Sabater}, {Scaife}, {Scheers},
  {Sendlinger}, {Shulevski}, {Sipior}, {Sobey}, {Stewart}, {Stroe}, {Swinbank},
  {Tasse}, {Tr{\"u}stedt}, {Varenius}, {van Velzen}, {Vilchez}, {van Weeren},
  {Wijnholds}, {Williams}, {de Bruyn}, {Nijboer}, {Wise}, {Alexov}, {Anderson},
  {Avruch}, {Beck}, {Bell}, {van Bemmel}, {Bentum}, {Bernardi}, {Best},
  {Breitling}, {Brouw}, {Br{\"u}ggen}, {Butcher}, {Ciardi}, {Conway}, {de
  Geus}, {de Jong}, {de Vos}, {Deller}, {Dettmar}, {Duscha}, {Eisl{\"o}ffel},
  {Engels}, {Falcke}, {Fender}, {Garrett}, {Grie{\ss}meier}, {Gunst},
  {Hamaker}, {Hessels}, {Hoeft}, {H{\"o}randel}, {Holties}, {Intema},
  {Jackson}, {J{\"u}tte}, {Karastergiou}, {Klijn}, {Kondratiev}, {Koopmans},
  {Kuniyoshi}, {Kuper}, {Law}, {van Leeuwen}, {Loose}, {Maat}, {Markoff},
  {McFadden}, {McKay-Bukowski}, {Mevius}, {Miller-Jones}, {Morganti}, {Munk},
  {Nelles}, {Noordam}, {Norden}, {Paas}, {Polatidis}, {Reich}, {Renting},
  {R{\"o}ttgering}, {Schoenmakers}, {Schwarz}, {Sluman}, {Smirnov}, {Stappers},
  {Steinmetz}, {Tagger}, {Tang}, {ter Veen}, {Thoudam}, {Vermeulen}, {Vocks},
  {Vogt}, {Wijers}, {Wucknitz}, {Yatawatta}, \& {Zarka}}]{2015A&A...582A.123H}
{Heald}, G.~H. {et~al.} 2015, \aap, 582, A123

\bibitem[{{Hinshaw} {et~al.}(2013){Hinshaw}, {Larson}, {Komatsu}, {Spergel},
  {Bennett}, {Dunkley}, {Nolta}, {Halpern}, {Hill}, {Odegard}, {Page}, {Smith},
  {Weiland}, {Gold}, {Jarosik}, {Kogut}, {Limon}, {Meyer}, {Tucker}, {Wollack},
  \& {Wright}}]{2013ApJS..208...19H_wmap9_parameters}
{Hinshaw}, G. {et~al.} 2013, \apjs, 208, 19

\bibitem[{{Hurley-Walker} {et~al.}(2014){Hurley-Walker}, {Morgan}, {Wayth},
  {Hancock}, {Bell}, {Bernardi}, {Bhat}, {Briggs}, {Deshpande}, {Ewall-Wice},
  {Feng}, {Hazelton}, {Hindson}, {Jacobs}, {Kaplan}, {Kudryavtseva}, {Lenc},
  {McKinley}, {Mitchell}, {Pindor}, {Procopio}, {Oberoi}, {Offringa}, {Ord},
  {Riding}, {Bowman}, {Cappallo}, {Corey}, {Emrich}, {Gaensler}, {Goeke},
  {Greenhill}, {Hewitt}, {Johnston-Hollitt}, {Kasper}, {Kratzenberg},
  {Lonsdale}, {Lynch}, {McWhirter}, {Morales}, {Morgan}, {Prabu}, {Rogers},
  {Roshi}, {Shankar}, {Srivani}, {Subrahmanyan}, {Tingay}, {Waterson},
  {Webster}, {Whitney}, {Williams}, \& {Williams}}]{Hurley-walker:2014p45}
{Hurley-Walker}, N. {et~al.} 2014, \pasa, 31, 45

\bibitem[{{Intema} {et~al.}(2016){Intema}, {Jagannathan}, {Mooley}, \&
  {Frail}}]{Intema:2016arXiv160304368I}
{Intema}, H.~T., {Jagannathan}, P., {Mooley}, K.~P., \& {Frail}, D.~A. 2016,
  ArXiv e-prints

\bibitem[{Jacobs {et~al.}(2011)Jacobs, Aguirre, Parsons, Pober, Bradley,
  Carilli, Gugliucci, Manley, Merwe, Moore, \& Parashare}]{Jacobs:2011p8438}
Jacobs, D.~C. {et~al.} 2011, The Astrophysical Journal, 734, L34

\bibitem[{{Jacobs} {et~al.}(2013){Jacobs}, {Parsons}, {Aguirre}, {Ali},
  {Bowman}, {Bradley}, {Carilli}, {DeBoer}, {Dexter}, {Gugliucci}, {Klima},
  {MacMahon}, {Manley}, {Moore}, {Pober}, {Stefan}, \&
  {Walbrugh}}]{2013ApJ...776..108J}
{Jacobs}, D.~C. {et~al.} 2013, \apj, 776, 108

\bibitem[{{Jacobs} {et~al.}(2015){Jacobs}, {Pober}, {Parsons}, {Aguirre},
  {Ali}, {Bowman}, {Bradley}, {Carilli}, {DeBoer}, {Dexter}, {Gugliucci},
  {Klima}, {Liu}, {MacMahon}, {Manley}, {Moore}, {Stefan}, \&
  {Walbrugh}}]{2015ApJ...801...51J}
---. 2015, \apj, 801, 51

\bibitem[{Jeli{\'c} {et~al.}(2008)Jeli{\'c}, Zaroubi, Labropoulos, Thomas,
  Bernardi, Brentjens, de~Bruyn, Ciardi, Harker, Koopmans, Pandey, Schaye, \&
  Yatawatta}]{Jelic:2008p2130}
Jeli{\'c}, V. {et~al.} 2008, Monthly Notices of the Royal Astronomical Society,
  389, 1319, (c) Journal compilation {\copyright} 2008 RAS

\bibitem[{{Koopmans} {et~al.}(2014){Koopmans}, {Pritchard}, {Mellema},
  {Aguirre}, {Ahn}, {Barkana}, {van Bemmel}, {Bernardi}, {Bonaldi}, {Briggs},
  {de Bruyn}, {Chang}, {Chapman}, {Chen}, {Ciardi}, {Dayal}, {Ferrara},
  {Fialkov}, {Fiore}, {Ichiki}, {Illiev}, {Inoue}, {Jelic}, {Jones}, {Lazio},
  {Maio}, {Majumdar}, {Mack}, {Mesinger}, {Morales}, {Parsons}, {Pen},
  {Santos}, {Schneider}, {Semelin}, {de Souza}, {Subrahmanyan}, {Takeuchi},
  {Vedantham}, {Wagg}, {Webster}, {Wyithe}, {Datta}, \&
  {Trott}}]{2014aska.confE...1K}
{Koopmans}, L. {et~al.} 2014, in Proceedings of Advancing Astrophysics with the
  Square Kilometre Array (AASKA14). 9 -13 June, 2014. Giardini Naxos, Italy.
  (http://pos.sissa.it/cgi-bin/reader/conf.cgi?confid=215), 1

\bibitem[{{Lane} {et~al.}(2014){Lane}, {Cotton}, {van Velzen}, {Clarke},
  {Kassim}, {Helmboldt}, {Lazio}, \& {Cohen}}]{2014MNRAS.440..327L}
{Lane}, W.~M., {Cotton}, W.~D., {van Velzen}, S., {Clarke}, T.~E., {Kassim},
  N.~E., {Helmboldt}, J.~F., {Lazio}, T.~J.~W., \& {Cohen}, A.~S. 2014, \mnras,
  440, 327

\bibitem[{Large {et~al.}(1991)Large, Cram, \& Burgess}]{Large:1991p7760}
Large, M., Cram, L., \& Burgess, A. 1991, The Observatory, 111, 72

\bibitem[{{Liu} {et~al.}(2014{\natexlab{a}}){Liu}, {Parsons}, \&
  {Trott}}]{2014PhRvD..90b3018L}
{Liu}, A., {Parsons}, A.~R., \& {Trott}, C.~M. 2014{\natexlab{a}}, \prd, 90,
  023018

\bibitem[{{Liu} {et~al.}(2014{\natexlab{b}}){Liu}, {Parsons}, \&
  {Trott}}]{2014PhRvD..90b3019L}
---. 2014{\natexlab{b}}, \prd, 90, 023019

\bibitem[{Liu \& Tegmark(2011)}]{Liu:2011p8763}
Liu, A. \& Tegmark, M. 2011, Physical Review D, 83, 103006

\bibitem[{Liu {et~al.}(2009)Liu, Tegmark, \& Zaldarriaga}]{Liu:2009p4762}
Liu, A., Tegmark, M., \& Zaldarriaga, M. 2009, MNRAS, 394, 1575, (c) Journal
  compilation {\copyright} 2009 RAS

\bibitem[{Lonsdale {et~al.}(2009)Lonsdale, Cappallo, Morales, Briggs,
  Benkevitch, Bowman, Bunton, Burns, Corey, Desouza, Doeleman, Derome,
  Deshpande, Gopala, Greenhill, Herne, Hewitt, Kamini, Kasper, Kincaid, Kocz,
  Kowald, Kratzenberg, Kumar, Lynch, Madhavi, Matejek, Mitchell, Morgan,
  Oberoi, Ord, Pathikulangara, Prabu, Rogers, Roshi, Salah, Sault, Shankar,
  Srivani, Stevens, Tingay, Vaccarella, Waterson, Wayth, Webster, Whitney,
  Williams, \& Williams}]{Lonsdale:2009p7913}
Lonsdale, C.~J. {et~al.} 2009, Proceedings of the IEEE, 97, 1497

\bibitem[{Mauch {et~al.}(2003)Mauch, Murphy, Buttery, Curran, Hunstead,
  Piestrzynski, Robertson, \& Sadler}]{Mauch:2003p8804}
Mauch, T., Murphy, T., Buttery, H.~J., Curran, J., Hunstead, R.~W.,
  Piestrzynski, B., Robertson, J.~G., \& Sadler, E.~M. 2003, Monthly Notice of
  the Royal Astronomical Society, 342, 1117

\bibitem[{{Mitchell} {et~al.}(2008){Mitchell}, {Greenhill}, {Wayth}, {Sault},
  {Lonsdale}, {Cappallo}, {Morales}, \& {Ord}}]{Mitchell:2008p707}
{Mitchell}, D.~A., {Greenhill}, L.~J., {Wayth}, R.~B., {Sault}, R.~J.,
  {Lonsdale}, C.~J., {Cappallo}, R.~J., {Morales}, M.~F., \& {Ord}, S.~M. 2008,
  IEEE Journal of Selected Topics in Signal Processing, 2, 707

\bibitem[{Morales {et~al.}(2006{\natexlab{a}})Morales, Bowman, Cappallo, \&
  Hewitt}]{Morales:2006p1870}
Morales, M., Bowman, J., Cappallo, R., \& Hewitt, J. 2006{\natexlab{a}}, New
  Astronomy Reviews

\bibitem[{Morales {et~al.}(2006{\natexlab{b}})Morales, Bowman, \&
  Hewitt}]{Morales:2006p1903}
Morales, M., Bowman, J., \& Hewitt, J. 2006{\natexlab{b}}, The Astrophysical
  Journal

\bibitem[{Morales {et~al.}(2012)Morales, Hazelton, Sullivan, \&
  Beardsley}]{Morales:2012p8790}
Morales, M.~F., Hazelton, B., Sullivan, I., \& Beardsley, A. 2012, ApJ, 752,
  137

\bibitem[{Morales \& Wyithe(2010)}]{Morales:2010p8093}
Morales, M.~F. \& Wyithe, J. S.~B. 2010, Annual review of astronomy and
  astrophysics, 48, 127, oise

\bibitem[{{Neben} {et~al.}(2015){Neben}, {Bradley}, {Hewitt}, {Bernardi},
  {Bowman}, {Briggs}, {Cappallo}, {Deshpande}, {Goeke}, {Greenhill},
  {Hazelton}, {Johnston-Hollitt}, {Kaplan}, {Lonsdale}, {McWhirter},
  {Mitchell}, {Morales}, {Morgan}, {Oberoi}, {Ord}, {Prabu}, {Shankar},
  {Srivani}, {Subrahmanyan}, {Tingay}, {Wayth}, {Webster}, {Williams}, \&
  {Williams}}]{2015RaSc...50..614N}
{Neben}, A.~R. {et~al.} 2015, Radio Science, 50, 614

\bibitem[{Noordam(2004)}]{Noordam:2004p2379}
Noordam, J.~E. 2004, Ground-based Telescopes. Edited by Oschmann, 5489, 817

\bibitem[{{Offringa} {et~al.}(2010){Offringa}, {de Bruyn}, {Zaroubi}, \&
  {Biehl}}]{offringa:2010rfim.workE..36O}
{Offringa}, A.~R., {de Bruyn}, A.~G., {Zaroubi}, S., \& {Biehl}, M. 2010, in
  RFI Mitigation Workshop, 36

\bibitem[{{Offringa} {et~al.}(2014){Offringa}, {McKinley}, {Hurley-Walker},
  {Briggs}, {Wayth}, {Kaplan}, {Bell}, {Feng}, {Neben}, {Hughes}, {Rhee},
  {Murphy}, {Bhat}, {Bernardi}, {Bowman}, {Cappallo}, {Corey}, {Deshpande},
  {Emrich}, {Ewall-Wice}, {Gaensler}, {Goeke}, {Greenhill}, {Hazelton},
  {Hindson}, {Johnston-Hollitt}, {Jacobs}, {Kasper}, {Kratzenberg}, {Lenc},
  {Lonsdale}, {Lynch}, {McWhirter}, {Mitchell}, {Morales}, {Morgan},
  {Kudryavtseva}, {Oberoi}, {Ord}, {Pindor}, {Procopio}, {Prabu}, {Riding},
  {Roshi}, {Shankar}, {Srivani}, {Subrahmanyan}, {Tingay}, {Waterson},
  {Webster}, {Whitney}, {Williams}, \& {Williams}}]{2014MNRAS.444..606O}
{Offringa}, A.~R. {et~al.} 2014, \mnras, 444, 606

\bibitem[{{Offringa} {et~al.}(2015){Offringa}, {Wayth}, {Hurley-Walker},
  {Kaplan}, {Barry}, {Beardsley}, {Bell}, {Bernardi}, {Bowman}, {Briggs},
  {Callingham}, {Cappallo}, {Carroll}, {Deshpande}, {Dillon}, {Dwarakanath},
  {Ewall-Wice}, {Feng}, {For}, {Gaensler}, {Greenhill}, {Hancock}, {Hazelton},
  {Hewitt}, {Hindson}, {Jacobs}, {Johnston-Hollitt}, {Kapi{\'n}ska}, {Kim},
  {Kittiwisit}, {Lenc}, {Line}, {Loeb}, {Lonsdale}, {McKinley}, {McWhirter},
  {Mitchell}, {Morales}, {Morgan}, {Morgan}, {Neben}, {Oberoi}, {Ord}, {Paul},
  {Pindor}, {Pober}, {Prabu}, {Procopio}, {Riding}, {Udaya Shankar}, {Sethi},
  {Srivani}, {Staveley-Smith}, {Subrahmanyan}, {Sullivan}, {Tegmark},
  {Thyagarajan}, {Tingay}, {Trott}, {Webster}, {Williams}, {Williams}, {Wu},
  {Wyithe}, \& {Zheng}}]{2015PASA...32....8O}
---. 2015, \pasa, 32, 8

\bibitem[{{Ord} {et~al.}(2015){Ord}, {Crosse}, {Emrich}, {Pallot}, {Wayth},
  {Clark}, {Tremblay}, {Arcus}, {Barnes}, {Bell}, {Bernardi}, {Bhat}, {Bowman},
  {Briggs}, {Bunton}, {Cappallo}, {Corey}, {Deshpande}, {deSouza},
  {Ewell-Wice}, {Feng}, {Goeke}, {Greenhill}, {Hazelton}, {Herne}, {Hewitt},
  {Hindson}, {Hurley-Walker}, {Jacobs}, {Johnston-Hollitt}, {Kaplan}, {Kasper},
  {Kincaid}, {Koenig}, {Kratzenberg}, {Kudryavtseva}, {Lenc}, {Lonsdale},
  {Lynch}, {McKinley}, {McWhirter}, {Mitchell}, {Morales}, {Morgan}, {Oberoi},
  {Offringa}, {Pathikulangara}, {Pindor}, {Prabu}, {Procopio}, {Remillard},
  {Riding}, {Rogers}, {Roshi}, {Salah}, {Sault}, {Udaya Shankar}, {Srivani},
  {Stevens}, {Subrahmanyan}, {Tingay}, {Waterson}, {Webster}, {Whitney},
  {Williams}, {Williams}, \& {Wyithe}}]{Ord:2015PASA...32....6O}
{Ord}, S.~M. {et~al.} 2015, \pasa, 32, 6

\bibitem[{Ord {et~al.}(2010)Ord, Mitchell, Wayth, Greenhill, Bernardi, Gleadow,
  Edgar, Clark, Allen, Arcus, Benkevitch, Bowman, Briggs, Bunton, Burns,
  Cappallo, Coles, Corey, Desouza, Doeleman, Derome, Deshpande, Emrich, Goeke,
  Gopalakrishna, Herne, Hewitt, Kamini, Kaplan, Kasper, Kincaid, Kocz, Kowald,
  Kratzenberg, Kumar, Lonsdale, Lynch, McWhirter, Madhavi, Matejek, Morales,
  Morgan, Oberoi, Pathikulangara, Prabu, Rogers, Roshi, Salah, Schinkel,
  Shankar, Srivani, Stevens, Tingay, Vaccarella, Waterson, Webster, Whitney,
  Williams, \& Williams}]{Ord:2010p8442}
Ord, S.~M. {et~al.} 2010, Publications of the Astronomical Society of the
  Pacific, 122, 1353

\bibitem[{Parsons {et~al.}(2014)Parsons, Liu, Aguirre, Ali, Bradley, Carilli,
  DeBoer, Dexter, Gugliucci, Jacobs, Klima, Macmahon, Manley, Moore, Pober,
  Stefan, \& Walbrugh}]{Parsons:2014p10499}
Parsons, A.~R. {et~al.} 2014, ApJ, 788, 106

\bibitem[{Parsons {et~al.}(2012)Parsons, Pober, Aguirre, Carilli, Jacobs, \&
  Moore}]{Parsons:2012p8896}
Parsons, A.~R., Pober, J.~C., Aguirre, J.~E., Carilli, C.~L., Jacobs, D.~C., \&
  Moore, D.~F. 2012, The Astrophysical Journal, 756, 165

\bibitem[{Pindor {et~al.}(2011)Pindor, Wyithe, Mitchell, Ord, Wayth, \&
  Greenhill}]{Pindor:2011p10350}
Pindor, B., Wyithe, J. S.~B., Mitchell, D.~A., Ord, S.~M., Wayth, R.~B., \&
  Greenhill, L.~J. 2011, Publications of the Astronomical Society of Australia,
  28, 46, (c) 2011 Astronomical Society of Australia

\bibitem[{{Pober} {et~al.}(2016){Pober}, {Hazelton}, {Beardsley}, {Barry},
  {Martinot}, {Sullivan}, {Morales}, {Bell}, {Bernardi}, {Bhat}, {Bowman},
  {Briggs}, {Cappallo}, {Carroll}, {Corey}, {de Oliveira-Costa}, {Deshpande},
  {Dillon}, {Emrich}, {Ewall-Wice}, {Feng}, {Goeke}, {Greenhill}, {Hewitt},
  {Hindson}, {Hurley-Walker}, {Jacobs}, {Johnston-Hollitt}, {Kaplan}, {Kasper},
  {Kim}, {Kittiwisit}, {Kratzenberg}, {Kudryavtseva}, {Lenc}, {Line}, {Loeb},
  {Lonsdale}, {Lynch}, {McKinley}, {McWhirter}, {Mitchell}, {Morgan}, {Neben},
  {Oberoi}, {Offringa}, {Ord}, {Paul}, {Pindor}, {Prabu}, {Procopio}, {Riding},
  {Rogers}, {Roshi}, {Sethi}, {Udaya Shankar}, {Srivani}, {Subrahmanyan},
  {Tegmark}, {Thyagarajan}, {Tingay}, {Trott}, {Waterson}, {Wayth}, {Webster},
  {Whitney}, {Williams}, {Williams}, \& {Wyithe}}]{Pober:2016ApJ...819....8P}
{Pober}, J.~C. {et~al.} 2016, \apj, 819, 8

\bibitem[{Pober {et~al.}(2014)Pober, Liu, Dillon, Aguirre, Bowman, Bradley,
  Carilli, DeBoer, Hewitt, Jacobs, McQuinn, Morales, Parsons, Tegmark, \&
  Werthimer}]{Pober:2014p10390}
Pober, J.~C. {et~al.} 2014, The Astrophysical Journal, 782, 66

\bibitem[{Pober {et~al.}(2013)Pober, Parsons, Aguirre, Ali, Bradley, Carilli,
  DeBoer, Dexter, Gugliucci, Jacobs, Klima, MacMahon, Manley, Moore, Stefan, \&
  Walbrugh}]{Pober:2013p9942}
---. 2013, The Astrophysical Journal, 768, L36

\bibitem[{Pritchard \& Loeb(2012)}]{Pritchard:2012p9555}
Pritchard, J.~R. \& Loeb, A. 2012, Reports on Progress in Physics, 75, 6901

\bibitem[{{Salvini} \& {Wijnholds}(2014)}]{sal14}
{Salvini}, S. \& {Wijnholds}, S.~J. 2014, \aap, 571, A97

\bibitem[{Santos \& Cooray(2006)}]{Santos:2006p6697}
Santos, M.~G. \& Cooray, A. 2006, Physical Review D, 74, 83517

\bibitem[{Sullivan {et~al.}(2012)Sullivan, Morales, Hazelton, Arcus, Barnes,
  Bernardi, Briggs, Bowman, Bunton, Cappallo, Corey, Deshpande, Desouza,
  Emrich, Gaensler, Goeke, Greenhill, Herne, Hewitt, Johnston-Hollitt, Kaplan,
  Kasper, Kincaid, Koenig, Kratzenberg, Lonsdale, Lynch, McWhirter, Mitchell,
  Morgan, Oberoi, Ord, Pathikulangara, Prabu, Remillard, Rogers, Roshi, Salah,
  Sault, Shankar, Srivani, Stevens, Subrahmanyan, Tingay, Wayth, Waterson,
  Webster, Whitney, Williams, Williams, \& Wyithe}]{Sullivan:2012p9457}
Sullivan, I.~S. {et~al.} 2012, The Astrophysical Journal, 759, 17

\bibitem[{{Sutinjo} {et~al.}(2015){Sutinjo}, {O'Sullivan}, {Lenc}, {Wayth},
  {Padhi}, {Hall}, \& {Tingay}}]{Sutinjo:2015RaSc...50...52S}
{Sutinjo}, A., {O'Sullivan}, J., {Lenc}, E., {Wayth}, R.~B., {Padhi}, S.,
  {Hall}, P., \& {Tingay}, S.~J. 2015, Radio Science, 50, 52

\bibitem[{Tasse {et~al.}(2012)Tasse, van Diepen, van~der Tol, van Weeren, van
  Zwieten, Batejat, Bhatnagar, van Bemmel, B{\^\i}rzan, Bonafede, Conway,
  Ferrari, de~Gasperin, Golap, Heald, Jackson, Macario, McKean, Mohan,
  Orr{\`u}, Pizzo, Rafferty, Rau, R{\"o}ttgering, Shulevski, \&
  Collaboration}]{Tasse:2012p9459}
Tasse, C. {et~al.} 2012, Comptes Rendus Physique, 13, 28, acad{\'e}mie des
  sciences

\bibitem[{{Tegmark}(1997)}]{1997ApJ...480L..87T}
{Tegmark}, M. 1997, \apjl, 480, L87

\bibitem[{{Thompson} {et~al.}(2007){Thompson}, {Moran}, \& {Swenson}}]{2007TMS}
{Thompson}, A.~R., {Moran}, J.~M., \& {Swenson}, G.~W. 2007, {Interferometry
  and Synthesis in Radio Astronomy, .} (John Wiley \& Sons, 2007)

\bibitem[{{Thyagarajan} {et~al.}(2015{\natexlab{a}}){Thyagarajan}, {Jacobs},
  {Bowman}, {Barry}, {Beardsley}, {Bernardi}, {Briggs}, {Cappallo}, {Carroll},
  {Corey}, {de Oliveira-Costa}, {Dillon}, {Emrich}, {Ewall-Wice}, {Feng},
  {Goeke}, {Greenhill}, {Hazelton}, {Hewitt}, {Hurley-Walker},
  {Johnston-Hollitt}, {Kaplan}, {Kasper}, {Kim}, {Kittiwisit}, {Kratzenberg},
  {Lenc}, {Line}, {Loeb}, {Lonsdale}, {Lynch}, {McKinley}, {McWhirter},
  {Mitchell}, {Morales}, {Morgan}, {Neben}, {Oberoi}, {Offringa}, {Ord},
  {Paul}, {Pindor}, {Pober}, {Prabu}, {Procopio}, {Riding}, {Rogers}, {Roshi},
  {Udaya Shankar}, {Sethi}, {Srivani}, {Subrahmanyan}, {Sullivan}, {Tegmark},
  {Tingay}, {Trott}, {Waterson}, {Wayth}, {Webster}, {Whitney}, {Williams},
  {Williams}, {Wu}, \& {Wyithe}}]{2015ApJ...804...14T}
{Thyagarajan}, N. {et~al.} 2015{\natexlab{a}}, \apj, 804, 14

\bibitem[{{Thyagarajan} {et~al.}(2015{\natexlab{b}}){Thyagarajan}, {Jacobs},
  {Bowman}, {Barry}, {Beardsley}, {Bernardi}, {Briggs}, {Cappallo}, {Carroll},
  {Deshpande}, {de Oliveira-Costa}, {Dillon}, {Ewall-Wice}, {Feng},
  {Greenhill}, {Hazelton}, {Hernquist}, {Hewitt}, {Hurley-Walker},
  {Johnston-Hollitt}, {Kaplan}, {Kim}, {Kittiwisit}, {Lenc}, {Line}, {Loeb},
  {Lonsdale}, {McKinley}, {McWhirter}, {Mitchell}, {Morales}, {Morgan},
  {Neben}, {Oberoi}, {Offringa}, {Ord}, {Paul}, {Pindor}, {Pober}, {Prabu},
  {Procopio}, {Riding}, {Udaya Shankar}, {Sethi}, {Srivani}, {Subrahmanyan},
  {Sullivan}, {Tegmark}, {Tingay}, {Trott}, {Wayth}, {Webster}, {Williams},
  {Williams}, \& {Wyithe}}]{2015ApJ...807L..28T}
---. 2015{\natexlab{b}}, \apjl, 807, L28

\bibitem[{Thyagarajan {et~al.}(2013)Thyagarajan, Shankar, Subrahmanyan, Arcus,
  Bernardi, Bowman, Briggs, Bunton, Cappallo, Corey, deSouza, Emrich, Gaensler,
  Goeke, Greenhill, Hazelton, Herne, Hewitt, Johnston-Hollitt, Kaplan, Kasper,
  Kincaid, Koenig, Kratzenberg, Lonsdale, Lynch, Mcwhirter, Mitchell, Morales,
  Morgan, Oberoi, Ord, Pathikulangara, Remillard, Rogers, Roshi, Salah, Sault,
  Srivani, Stevens, Thiagaraj, Tingay, Wayth, Waterson, Webster, Whitney,
  Williams, Williams, \& Wyithe}]{Thyagarajan:2013p10039}
Thyagarajan, N. {et~al.} 2013, The Astrophysical Journal, 776, 6

\bibitem[{Tingay {et~al.}(2013)Tingay, Goeke, Bowman, Emrich, Ord, Mitchell,
  Morales, Booler, Crosse, Wayth, Lonsdale, Tremblay, Pallot, Colegate,
  Wicenec, Kudryavtseva, Arcus, Barnes, Bernardi, Briggs, Burns, Bunton,
  Cappallo, Corey, Deshpande, Desouza, Gaensler, Greenhill, Hall, Hazelton,
  Herne, Hewitt, Johnston-Hollitt, Kaplan, Kasper, Kincaid, Koenig,
  Kratzenberg, Lynch, Mckinley, Mcwhirter, Morgan, Oberoi, Pathikulangara,
  Prabu, Remillard, Rogers, Roshi, Salah, Sault, Udaya-Shankar, Schlagenhaufer,
  Srivani, Stevens, Subrahmanyan, Waterson, Webster, Whitney, Williams,
  Williams, \& Wyithe}]{Tingay:2013p9022}
Tingay, S. {et~al.} 2013, Publications of the Astronomical Society of
  Australia, 30, 7

\bibitem[{Trott {et~al.}(2012)Trott, Wayth, \& Tingay}]{Trott:2012p10466}
Trott, C., Wayth, R., \& Tingay, S. 2012, The Astrophysical Journal, 757, 101

\bibitem[{{Trott} {et~al.}(2016){Trott}, {Pindor}, {Procopio}, {Wayth},
  {Mitchell}, {McKinley}, {Tingay}, {Barry}, {Beardsley}, {Bernardi}, {Bowman},
  {Briggs}, {Cappallo}, {Carroll}, {de Oliveira-Costa}, {Dillon}, {Ewall-Wice},
  {Feng}, {Greenhill}, {Hazelton}, {Hewitt}, {Hurley-Walker},
  {Johnston-Hollitt}, {Jacobs}, {Kaplan}, {Kim}, {Lenc}, {Line}, {Loeb},
  {Lonsdale}, {Morales}, {Morgan}, {Neben}, {Thyagarajan}, {Oberoi},
  {Offringa}, {Ord}, {Paul}, {Pober}, {Prabu}, {Riding}, {Udaya Shankar},
  {Sethi}, {Srivani}, {Subrahmanyan}, {Sullivan}, {Tegmark}, {Webster},
  {Williams}, {Williams}, {Wu}, \& {Wyithe}}]{2016arXiv160102073T}
{Trott}, C.~M. {et~al.} 2016, ArXiv e-prints

\bibitem[{Vedantham {et~al.}(2012{\natexlab{a}})Vedantham, Shankar, \&
  Subrahmanyan}]{Vedantham:2012p10297}
Vedantham, H., Shankar, N.~U., \& Subrahmanyan, R. 2012{\natexlab{a}}, The
  Astrophysical Journal, 745, 176

\bibitem[{Vedantham {et~al.}(2012{\natexlab{b}})Vedantham, Shankar, \&
  Subrahmanyan}]{Vedantham:2012p9026}
---. 2012{\natexlab{b}}, The Astrophysical Journal, 745, 176

\bibitem[{{Wayth} {et~al.}(2015){Wayth}, {Lenc}, {Bell}, {Callingham},
  {Dwarakanath}, {Franzen}, {For}, {Gaensler}, {Hancock}, {Hindson},
  {Hurley-Walker}, {Jackson}, {Johnston-Hollitt}, {Kapi{\'n}ska}, {McKinley},
  {Morgan}, {Offringa}, {Procopio}, {Staveley-Smith}, {Wu}, {Zheng}, {Trott},
  {Bernardi}, {Bowman}, {Briggs}, {Cappallo}, {Corey}, {Deshpande}, {Emrich},
  {Goeke}, {Greenhill}, {Hazelton}, {Kaplan}, {Kasper}, {Kratzenberg},
  {Lonsdale}, {Lynch}, {McWhirter}, {Mitchell}, {Morales}, {Morgan}, {Oberoi},
  {Ord}, {Prabu}, {Rogers}, {Roshi}, {Shankar}, {Srivani}, {Subrahmanyan},
  {Tingay}, {Waterson}, {Webster}, {Whitney}, {Williams}, \&
  {Williams}}]{2015PASA...32...25W}
{Wayth}, R.~B. {et~al.} 2015, \pasa, 32, 25

\bibitem[{Yatawatta {et~al.}(2013)Yatawatta, de~Bruyn, Brentjens, Labropoulos,
  Pandey, Kazemi, Zaroubi, Koopmans, Offringa, Jeli{\'c}, Rubi, Veligatla,
  Wijnholds, Brouw, Bernardi, Ciardi, Daiboo, Harker, Mellema, Schaye, Thomas,
  Vedantham, Chapman, Abdalla, Alexov, Anderson, Avruch, Batejat, Bell, Bell,
  Bentum, Best, Bonafede, Bregman, Breitling, van~de Brink, Broderick,
  Br{\"u}ggen, Conway, de~Gasperin, de~Geus, Duscha, Falcke, Fallows, Ferrari,
  Frieswijk, Garrett, Griessmeier, Gunst, Hassall, Hessels, Hoeft, Iacobelli,
  Juette, Karastergiou, Kondratiev, Kramer, Kuniyoshi, Kuper, Leeuwen, Maat,
  Mann, McKean, Mevius, Mol, Munk, Nijboer, Noordam, Norden, Orru, Paas,
  Pandey-Pommier, Pizzo, Polatidis, Reich, R{\"o}ttgering, Sluman, Smirnov,
  Stappers, Steinmetz, Tagger, Tang, Tasse, ter Veen, Vermeulen, van Weeren,
  Wise, Wucknitz, \& Zarka}]{Yatawatta:2013p9699}
Yatawatta, S. {et~al.} 2013, Astronomy {\&} Astrophysics, 550, 136

\bibitem[{{Zaroubi}(2013)}]{zaroubi2013epoch}
{Zaroubi}, S. 2013, in Astrophysics and Space Science Library, Vol. 396,
  Astrophysics and Space Science Library, ed. T.~{Wiklind}, B.~{Mobasher}, \&
  V.~{Bromm}, 45

\end{thebibliography}
